\documentclass [10pt,eqsecnum,amsfonts,aps]{revtex4}

\input epsf
\newcommand{\beq}{\begin{equation}}
\newcommand{\eeq}{\end{equation}}
\newcommand{\beqs}{\begin{eqnarray}}
\newcommand{\eeqs}{\end{eqnarray}}

\newtheorem{theo}{Theorem}[section]

\begin{document}
\baselineskip 5.0mm

\title{Structure of the Partition Function and Transfer Matrices for the 
Potts Model in a Magnetic Field on Lattice Strips}

\bigskip

\author{Shu-Chiuan Chang$^{a}$}
\email{scchang@mail.ncku.edu.tw}

\author{Robert Shrock$^{b}$}
\email{robert.shrock@stonybrook.edu}

\affiliation{(a) \ Department of Physics \\
National Cheng Kung University \\
Tainan 70101, Taiwan}

\affiliation{(b) \ C. N. Yang Institute for Theoretical Physics \\
State University of New York \\
Stony Brook, N. Y. 11794}


\begin{abstract}

We determine the general structure of the partition function of the $q$-state
Potts model in an external magnetic field, $Z(G,q,v,w)$ for arbitrary $q$,
temperature variable $v$, and magnetic field variable $w$, on cyclic, M\"obius,
and free strip graphs $G$ of the square (sq), triangular (tri), and honeycomb
(hc) lattices with width $L_y$ and arbitrarily great length $L_x$. 
For the cyclic case we prove that the partition function has the form
$Z(\Lambda,L_y \times L_x,q,v,w)=\sum_{d=0}^{L_y} \tilde c^{(d)}
Tr[(T_{Z,\Lambda,L_y,d})^m]$, where $\Lambda$ denotes the lattice type, $\tilde
c^{(d)}$ are specified polynomials of degree $d$ in $q$, $T_{Z,\Lambda,L_y,d}$
is the corresponding transfer matrix, and $m=L_x$ ($L_x/2$) for $\Lambda=sq, \
tri \ (hc)$, respectively. An analogous formula is given for M\"obius strips,
while only $T_{Z,\Lambda,L_y,d=0}$ appears for free strips.  We exhibit a
method for calculating $T_{Z,\Lambda,L_y,d}$ for arbitrary $L_y$ and give
illustrative examples.  Explicit results for arbitrary $L_y$ are presented for
$T_{Z,\Lambda,L_y,d}$ with $d=L_y$ and $d=L_y-1$.  We find very simple formulas
for the determinant $det(T_{Z,\Lambda,L_y,d})$.  We also give results for 
self-dual cyclic strips of the square lattice.

\end{abstract}

\maketitle


\pagestyle{plain}
\pagenumbering{arabic}

\section{Introduction}

The $q$-state Potts model has served as a valuable system for the study of
phase transitions and critical phenomena \cite{potts}-\cite{wubook}, and
recently there has been considerable interest in its connections with
mathematical graph theory \cite{welsh}-\cite{jemrev}. For two-dimensional
lattices, additional insights into the critical behavior have been obtained
from conformal algebra methods \cite{cardy,difran}. On a lattice, or, more
generally, on a graph $G$, at temperature $T$ and in an external magnetic field
$H$, the original Hamiltonian formulation of this model is defined by the
partition function
\beq
Z = \sum_{ \{ \sigma_n \} } e^{-\beta {\cal H}}
\label{z}
\eeq
with the Hamiltonian
\beq
{\cal H} = -J \sum_{\langle i j \rangle} \delta_{\sigma_i, \sigma_j} 
- H \sum_i \delta_{\sigma_i, 1} 
\label{ham}
\eeq
where $\sigma_i=1,...,q$ are classical spin variables on each vertex (site) $i
\in G$, $\beta = (k_BT)^{-1}$, and $\langle i j \rangle$ denote pairs of
adjacent vertices. The graph $G=G(V,E)$ is defined by its vertex (site) set $V$
and its edge (bond) set $E$; we denote the number of vertices of $G$ as
$n=n(G)$ and the number of edges of $G$ as $e(G)$. Without loss of generality,
we take $G$ to be connected and we take the magnetic field to single out the
spin value $\sigma_i=1$. We use the notation
\beq
K = \beta J \ , \quad h = \beta H \ , \quad y = e^K \ , \quad v = y-1 \ , 
\quad w=e^h
\label{kdef}
\eeq
Thus, the physical ranges of $v$ are $v \ge 0$ for the Potts ferromagnet, and
$-1 \le v \le 0$ for Potts antiferromagnet. Positive $H$ gives a weighting
that favors spin configurations in which spins have the value 1, while
negative $H$ disfavors such configurations.  For positive and negative $H$, the
physical range of $w$ is $w > 1$ and $0 \le w < 1$, respectively. 

The original definition of the Potts model, (\ref{z}) and (\ref{ham}),
requires $q$ to be in the set of positive integers ${\mathbb N}_+$.  This
restriction is removed for the zero-field Potts model by the Fortuin-Kasteleyn
cluster representation \cite{fk}
\beq
Z(G,q,v) = \sum_{G' \subseteq G} v^{e(G')}q^{k(G')}
\label{cluster}
\eeq
where $G$ is an arbitrary graph, $G'=(V,E')$ with $E' \subseteq E$ is a
spanning subgraph of $G$, and $k(G')$ denotes the number of (connected)
components of $G'$. Because (\ref{cluster}) does not contain any explicit
reference to the spins $\{\sigma_i\}$ or summation over spin configurations, it
allows one to define the zero-field Potts model partition function with $q$ not
necessarily restricted to the positive integers, ${\mathbb N}_+$.  The
zero-field Potts model partition function is equivalent to the Tutte (also
called Tutte-Whitney) polynomial $T(G,x,y)$, a function of major importance in
mathematical graph theory \cite{welsh}-\cite{jemrev}, \cite{tutte}-\cite{boll}
defined by
\beq
T(G,x,y) = \sum_{G' \subseteq G} (x-1)^{k(G')-k(G)}(y-1)^{c(G')} \ ,
\label{t}
\eeq
where $c(G')=e(G')+k(G')-n(G')$ is the number of independent cycles on $G'$.
The equivalence is
\beq
Z(G,q,v) = (x-1)^{k(G)}(y-1)^{n(G)}T(G,x,y) \ ,
\label{zt}
\eeq
where $x=1+(q/v)$, so $q=(x-1)(y-1)$.

In order to treat the Potts model in a magnetic field for non-integral $q$, it
is necessary to have a generalization of Eq. (\ref{cluster}), that is, a
formula for the partition function that does not make any explicit reference to
the spins or any summation over the spin values, since these spin values are
restricted to lie in ${\mathbb N}_+$.  F. Y. Wu succeeded in constructing such
a generalization, which, for an arbitrary graph $G$, expresses $Z(G,q,v,w)$ as
a sum of terms from spanning subgraphs $G'$ of $G$ \cite{wu78} (see also
\cite{wurev,wubook}).  Let us label each of the connected components of $G'$ as
$G'_i$, $i=1,...,k(G')$.  Wu's result is \cite{wurev,wubook,wu78}
\beq
Z(G,q,v,w) = \sum_{G' \subseteq G} v^{e(G')} \
\prod_{i=1}^{k(G')} \Big ( q-1 + w^{n(G'_i)} \Big )
\label{clusterw}
\eeq
Clearly, this formula defines $Z(G,q,v,w)$ in a manner such that $q$ need not
be in ${\mathbb N}_+$.  Eq. (\ref{clusterw}) also shows that $Z(G,q,v,w)$ is a
polynomial in the variables $q$, $v$, and $w$. In the limit $h \to -\infty$
(i.e., $w \to 0$), Eqs. (\ref{z}) and (\ref{ham}) show that configurations in
which any $\sigma_i=1$ make no contribution to $Z$, so that the model reduces
to the zero-field case with $q$ replaced by $q-1$:
\beq
Z(G,q,v,0) = Z(G,q-1,v,1) 
\label{zw10}
\eeq

In this paper we present a method for calculating transfer matrices for the
$q$-state Potts model partition functions $Z(G,q,v,w)$ in an external magnetic
field $H$, for arbitrary $q$ and temperature variable $v$, on cyclic, M\"obius
and free strip graphs $G$ of the square (sq), triangular (tri), and honeycomb
(hc) lattices with width $L_y$ vertices and with arbitrarily great length $L_x$
vertices.  Since this method enables one to calculate $Z(G,q,v,w)$ for
arbitrarily great strip lengths, it complements calculations for $L_x \times
L_y$ lattice patches based on enumeration of states (e.g., \cite{kc}).  Using
our transfer matrix method, we determine the general structure of this
partition function as a sum of powers of the eigenvalues of the transfer
matrix, multiplied by certain coefficients that depend only on $q$, not on $v$
or $w$.  The result that we find exhibits some interesting differences with the
form that has been established for the Potts model partition function on
lattice strips in the case of zero external field, and we explain how our more
general structure reduces to the zero-field form when the external field
vanishes.  We shall present explicit results for arbitrary $L_y$ given for
$T_{Z,\Lambda,L_y,d}$ with $d=L_y$ and $d=L_y-1$, and the determinant
$det(T_{Z,\Lambda,L_y,d})$. We have calculated the full transfer matrices up to
widths $L_y=3$ for the square, triangular, and honeycomb lattices and $L_y=2$
for the cyclic self-dual strip of the square lattice.  Since the total
dimensions of these transfer matrices increase very rapidly with strip width,
it is not feasible to present many of the explicit results here; instead, we
concentrate on general methods and results that hold for arbitrary $L_y$.  In
Ref. \cite{hl}, besides mentioning briefly our structural results for cyclic
lattice strips, we have used the Wu formula (\ref{clusterw}) to derive
properties of $Z(G,q,v,w)$, for arbitrary graphs $G$, concerning factorization,
monotonicity, and zero-free regions.  In Ref. \cite{hl} we have also presented
a generalization of the Tutte polynomial that corresponds to $Z(G,q,v,w)$ and
have formulated and discussed two related weighted graph coloring
problems. Some earlier work using transfer matrices for the calculations of the
zero-field Potts model partition function for arbitrary $q$ and $v$ on lattice
strips of fixed width and arbitrary length is in Refs. \cite{bn}-\cite{zt}.
Transfer matrix and related linear algebraic methods, as well as related
generating function methods, have also been used to calculate a particular
special case in zero field, namely the chromatic polynomial
\cite{biggs,boll,rtrev}; references to the literature can be found in reviews
such as Refs. \cite{wurev}-\cite{jemrev}.

\section{General Structure of Potts Model Partition Function on Lattice Strips
  in a Magnetic Field}

\subsection{Basic Method of Analysis and Structure for Cyclic and M\"obius 
Lattice Strips} 

In this section we derive the general structural form of the Potts model
partition function $Z(G_s,q,v,w)$ in an external magnetic field $H$ on lattice
strip graphs $G_s$.  We label the lattice type as $\Lambda$ and abbreviate the
three respective types as $sq$, $tri$, and $hc$.  Each strip involves a
longitudinal repetition of $m$ copies of a particular subgraph.  For the
square-lattice strips, this is a column of squares.  It is convenient to
represent the strip of the triangular lattice as obtained from the
corresponding strip of the square lattice via the insertion of diagonal edges
connecting, say, the upper-left to lower-right vertices in each square.  In
both of these cases, the length is $L_x=m$ vertices.  We represent the strip of
the honeycomb lattice in the form of bricks oriented horizontally.  In this
case, since there are two vertices in 1-1 correspondence with each horizontal
side of a brick, $L_x=2m$ vertices.  Summarizing for all of three lattices, the
relation between the number of vertices and the number of repeated copies is
\beq
L_x = \cases{ m & if $\Lambda=sq$ \ or \ $tri$ \ or $G_D$ \cr
              2m & if $\Lambda=hc$ \cr }
\label{lxm}
\eeq
Here $G_D$ is the cyclic self-dual strip of the square lattice, to be discussed
further below.

For cyclic strips, the full transfer matrix $T_{Z,\Lambda,L_y}$, has a block 
structure formally specified by
\beq
T_{Z,\Lambda,L_y} = \bigoplus_{d=0}^{L_y} \prod T_{Z,\Lambda,L_y,d} 
\label{Tdirectsum}
\eeq
where the product $\prod T_{Z,\Lambda,L_y,d}$ means a set of square blocks of
the form $\lambda_{Z,\Lambda,L_y,d,j}$ times the identity matrix.  As
indicated, each block is indexed by a non-negative integer $d$, which runs from
0 to $L_y$.  We shall refer to this as the degree of the block.  The reason
for this terminology is that for $q \ge 5$, this abstract submatrix has a
dimension given by a certain polynomial $\tilde c^{(d)}$ defined below in
Eq. (\ref{ctd}), which is of degree $d$ in $q$.  We shall also refer to the
$\lambda_{Z,\Lambda,L_y,d}$ in this block as lying in the degree-$d$
subspace of the full space in which the transfer matrix is defined. From
Eq. (\ref{Tdirectsum}), it follows that the partition function of the Potts
model, in an external magnetic field, on an $L_x \times L_y$ strip of the
lattice $\Lambda$ has the general structural form
\beqs
Z(\Lambda, L_y \times L_x,cyc.,q,v,w) 
& = & \sum_{d=0}^{L_y} \tilde c^{(d)} Tr[(T_{Z,\Lambda,L_y,d})^m] \cr\cr
& = & \sum_{d=0}^{L_y} \tilde c^{(d)} \sum_{j=1}^{n_{Zh}(\Lambda,L_y,d)} 
(\lambda_{Z,\Lambda,L_y,d,j})^m
\label{zgsum_transfer}
\eeqs
with $m$ given by (\ref{lxm}).  Here the eigenvalues
$\lambda_{Z,\Lambda,L_y,d,j}$ depend on the lattice type $\Lambda$, the strip
width $L_y$, and the variables $q$, $v$, and $w$, but not on the strip length,
$L_x$.  The number of different $\lambda_{Z,\Lambda,L_y,d}$'s in each subspace
of degree $d$ is given by $n_{Zh}(L_y,d)$, where we use the symbol $Zh$ to
indicate the nonzero field and to distinguish these numbers from the different
numbers $n_Z(L_y,d)$ for the zero-field case. The coefficients $\tilde c^{(d)}
\equiv \tilde c^{(d)}(q)$ are polynomials of degree $d$ in $q$ defined by:
\beq
\tilde c^{(d)} = 
\sum_{j=0}^d (-1)^j {2d-j \choose j}(q-1)^{d-j}
\label{ctd}
\eeq
The first few
of these coefficients are $\tilde c^{(0)} = 1$, $\tilde c^{(1)} = q-2$, $\tilde
c^{(2)} = q^2-5q+5$, $\tilde c^{(3)} = q^3-8q^2+19q-13$, and $\tilde c^{(4)} =
(q-2)(q^3-9q^2+24q-17)$.  For $q \ge 5$, the coefficients $\tilde
c^{(d)}$ are positive integers and can be interpreted as multiplicities of the
corresponding eigenvalues $\lambda_{Z,\Lambda,L_y,d}$, or equivalently, as the
dimensions of the sub-blocks $T_{Z,\Lambda,L_y,d}$ in the transfer matrix
$T_{Z,\Lambda,L_y}$.  Starting from this range of $q$, one can then continue
the expression (\ref{zgsum_transfer}) to arbitrary $q$.  However, for some
positive integer $q$ values, the $\tilde c^{(d)}$'s are negative, and hence
cannot directly be interpreted as multiplicities of eigenvalues. For example,
\beq
\tilde c^{(d)}=-1 \quad {\rm for} \quad q=2 \quad {\rm and} \quad 
d=2 \ {\rm mod} \ 3
\label{ctdm1a}
\eeq
and
\beq
\tilde c^{(d)}=-1 \quad {\rm for} \quad q=3 \quad {\rm and} \quad 
\quad d=2 \ {\rm or} \ 3 \ {\rm mod} \ 4
\label{ctdm1b}
\eeq
For brevity, we usually suppress the argument in the notation, writing simply
$\tilde c^{(d)}$ rather than $\tilde c^{(d)}(q)$.  The coefficients $\tilde
c^{(d)}$ for the present case of the Potts model in a nonzero magnetic field
are related to the corresponding coefficients $c^{(d)}$ \cite{saleur,cf} for
the zero-field Potts model according to the following equation (with arguments
indicated explicitly)
\beq
\tilde c^{(d)}(q) = c^{(d)}(q-1) \ , 
\label{ccrel}
\eeq
where
\beq
c^{(d)} = U_{2d}\Big (\frac{\sqrt{q}}{2} \Big ) = 
\sum_{j=0}^d (-1)^j {2d-j \choose j}q^{d-j}
\label{cd}
\eeq
where $U_n(x)$ is the Chebyshev polynomial of the second kind. We have also
constructed a general formula analogous to Eq. (\ref{zgsum_transfer}) for the
$Z(G,q,v,w)$ where $G$ is a M\"obius strip and a self-dual strip of the square
lattice, extending our earlier results for the zero-field case in
\cite{cf}-\cite{pt}.

To derive Eq. (\ref{zgsum_transfer}), let us first consider the subspace of
degree $d=0$ in the transfer matrix for a strip with width $L_y$ vertices. This
matrix is defined with respect to a given basis, and we shall refer to the
configurations that comprise this basis as the basis elements.  In addition to
the basis elements of the transfer matrix for the zero-field case consisting of
all the possible non-crossing partitions of $L_y$ vertices, there are
additional basis elements where certain vertices are in the $q=1$ state. The
eigenvalues of the transfer matrix for a free strip are the same as the
eigenvalues of the transfer matrix $T_{Z,\Lambda,L_y,d=0}$ in this degree $d=0$
subspace for the corresponding cyclic strip, where a set of horizontal edges
connecting two adjacent sets of $L_y$ vertices do not occur. The dimension of
this matrix, denoted as $n_{Zh}(L_y,0)$, is the binomial transform \cite{bs} of
a Catalan number,
\beq
n_{Zh}(L_y,0) = \sum_{k=0}^{L_y} {L_y \choose k} C_k 
\label{nzlyd0}
\eeq
where
\beq
C_k=\frac{1}{k+1}{2k \choose k}
\label{catalan}
\eeq
is the Catalan number. (No confusion should result from our use of the same
symbol $C_n$ to denote the circuit graph with $n$ vertices since the meaning
will be clear from context.) Parenthetically, we note that the binomial
transform in Eq. (\ref{nzlyd0}) has been of interest in other combinatorial
problems and appears as sequence A007317 in Ref. \cite{sl}; some other
sequences given here also have appeared in different mathematical contexts and
have been similarly catalogued in Ref. \cite{sl}.  We do not show the
dependence on the lattice $\Lambda$ explicitly in $n_{Zh}(L_y,d)$ as it is the
same for cyclic strips of the square, triangular and honeycomb lattices as for
the zero-field transfer matrices \cite{cf,hca}. (Below we shall consider
self-dual strips $G_D$ of the square lattice, which have different dimensions
$n_{Zh}(G_D,L_y,d)$; for this case we shall include the $G_D$ dependence in the
notation.)  The degree $d=1$ subspace is given by all of the possible
non-crossing partitions with a color assignment, out of $q-1$ states, to one
vertex, with possible connections with other vertices, plus the basis elements
where certain other vertices are in the $q=1$ state. The multiplicity is given
by $\tilde c^{(1)}=q-2$. This follows because there are $q-1$ possible ways of
making this color assignment, but one of these has to be subtracted, since the
effect of all the possible color assignments is equivalent to the choice of no
specific color assignment, which has been taken into account in the level 0
subspace. Equivalently, $\tilde c^{(1)}(q) = c^{(1)}(q-1)$.  In this derivation
and subsequent ones we assume that $q$ is an integer $\ge 5$ to begin with, so
that the multiplicities are positive-definite; we then analytically continue
them downward to apply in the region $0 \le q < 5$ where $\tilde c^{(d)}$ can
be zero or negative.  For the next subspace $d=2$ we consider all of the
non-crossing partitions with two-color assignments to two separated vertices
(with possible connections with other vertices), plus the basis elements where
certain other vertices are in the $q=1$ state.  This method is then continued
for higher $d$ up to the maximum degree, $d=L_y$. The multiplicity $\tilde
c^{(d)}(q)$ for general $d$ is given by Eq. (\ref{ctd}).

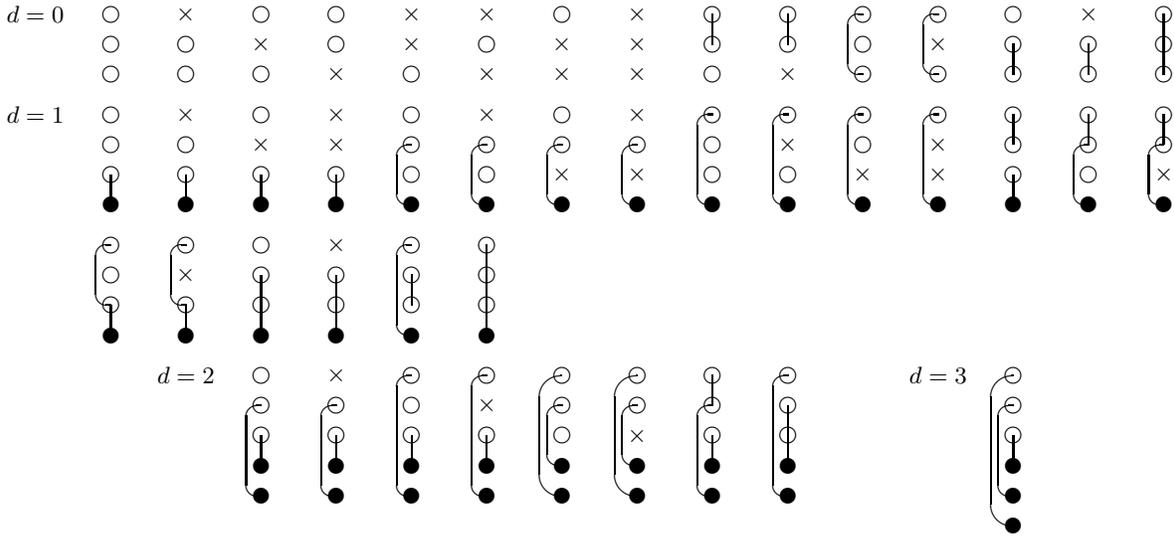
\begin{figure}
\unitlength 1mm 
\begin{picture}(150,8)
\put(0,8){\makebox(0,0){{\small $d=0$}}}
\multiput(10,0)(10,0){3}{\circle{2}}
\multiput(50,0)(40,0){2}{\circle{2}}
\multiput(110,0)(10,0){5}{\circle{2}}
\multiput(40,0)(60,0){2}{\makebox(0,0){$\times$}} 
\multiput(60,0)(10,0){3}{\makebox(0,0){$\times$}} 
\multiput(10,4)(10,0){2}{\circle{2}}
\multiput(40,4)(20,0){2}{\circle{2}}
\multiput(90,4)(10,0){3}{\circle{2}}
\multiput(130,4)(10,0){3}{\circle{2}}
\multiput(30,4)(20,0){3}{\makebox(0,0){$\times$}} 
\multiput(80,4)(40,0){2}{\makebox(0,0){$\times$}} 
\multiput(10,8)(20,0){2}{\circle{2}} 
\multiput(40,8)(30,0){2}{\circle{2}} 
\multiput(90,8)(10,0){5}{\circle{2}} 
\put(150,8){\circle{2}} 
\multiput(20,8)(30,0){3}{\makebox(0,0){$\times$}} 
\multiput(60,8)(80,0){2}{\makebox(0,0){$\times$}} 
\multiput(90,4)(10,0){2}{\line(0,1){4}}
\multiput(110,4)(10,0){2}{\oval(4,8)[l]} 
\multiput(130,0)(10,0){2}{\line(0,1){4}}
\put(150,0){\line(0,1){8}}
\end{picture}

\vspace*{5mm} 
\begin{picture}(150,12)
\put(0,12){\makebox(0,0){{\small $d=1$}}}
\multiput(10,0)(10,0){15}{\circle*{2}}
\multiput(10,4)(10,0){6}{\circle{2}}
\multiput(90,4)(10,0){2}{\circle{2}}
\multiput(130,4)(10,0){2}{\circle{2}}
\multiput(70,4)(10,0){2}{\makebox(0,0){$\times$}} 
\multiput(110,4)(10,0){2}{\makebox(0,0){$\times$}} 
\put(150,4){\makebox(0,0){$\times$}} 
\multiput(10,8)(10,0){2}{\circle{2}}
\multiput(50,8)(10,0){5}{\circle{2}}
\put(110,8){\circle{2}}
\multiput(130,8)(10,0){3}{\circle{2}}
\multiput(30,8)(10,0){2}{\makebox(0,0){$\times$}} 
\multiput(100,8)(20,0){2}{\makebox(0,0){$\times$}} 
\multiput(10,12)(20,0){4}{\circle{2}} 
\multiput(90,12)(10,0){7}{\circle{2}} 
\multiput(20,12)(20,0){4}{\makebox(0,0){$\times$}} 
\multiput(10,0)(10,0){4}{\line(0,1){4}}
\multiput(50,4)(10,0){4}{\oval(4,8)[l]} 
\multiput(90,6)(10,0){4}{\oval(4,12)[l]}
\multiput(130,0)(0,8){2}{\line(0,1){4}} 
\multiput(140,8)(10,0){2}{\line(0,1){4}}
\multiput(140,4)(10,0){2}{\oval(4,8)[l]} 
\end{picture}

\vspace*{5mm} 
\begin{picture}(150,12)
\multiput(10,0)(10,0){6}{\circle*{2}}
\multiput(10,4)(10,0){6}{\circle{2}}
\put(10,8){\circle{2}}
\multiput(30,8)(10,0){4}{\circle{2}}
\multiput(20,8)(20,4){2}{\makebox(0,0){$\times$}} 
\multiput(10,12)(10,0){3}{\circle{2}}
\multiput(50,12)(10,0){2}{\circle{2}}
\multiput(10,0)(10,0){2}{\line(0,1){4}} 
\multiput(10,8)(10,0){2}{\oval(4,8)[l]}
\multiput(30,0)(10,0){2}{\line(0,1){8}} 
\put(50,6){\oval(4,12)[l]}
\put(50,4){\line(0,1){4}} 
\put(60,0){\line(0,1){12}}
\end{picture}

\vspace*{5mm} 
\begin{picture}(110,20)
\put(0,20){\makebox(0,0){{\small $d=2$}}}
\multiput(10,4)(10,0){8}{\circle*{2}}
\multiput(10,8)(10,0){8}{\circle*{2}}
\multiput(10,12)(10,0){5}{\circle{2}}
\multiput(70,12)(10,0){2}{\circle{2}}
\put(60,12){\makebox(0,0){$\times$}} 
\multiput(10,16)(10,0){3}{\circle{2}}
\multiput(50,16)(10,0){4}{\circle{2}}
\put(40,16){\makebox(0,0){$\times$}} 
\put(10,20){\circle{2}} 
\multiput(30,20)(10,0){6}{\circle{2}} 
\put(20,20){\makebox(0,0){$\times$}} 
\multiput(10,8)(10,0){2}{\line(0,1){4}}
\multiput(10,10)(10,0){2}{\oval(4,12)[l]} 
\multiput(30,8)(10,0){2}{\line(0,1){4}}
\multiput(30,12)(10,0){2}{\oval(4,16)[l]} 
\multiput(50,12)(10,0){2}{\oval(4,8)[l]}
\multiput(50,12)(10,0){2}{\oval(6,16)[l]} 
\multiput(70,8)(0,8){2}{\line(0,1){4}}
\put(70,10){\oval(4,12)[l]} 
\put(80,8){\line(0,1){8}} 
\put(80,12){\oval(4,16)[l]}
\put(100,20){\makebox(0,0){{\small $d=3$}}}
\multiput(110,0)(0,4){3}{\circle*{2}}
\multiput(110,12)(0,4){3}{\circle{2}} 
\put(110,8){\line(0,1){4}}
\put(110,10){\oval(4,12)[l]} 
\put(110,10){\oval(6,20)[l]}
\end{picture}
\caption{\footnotesize{Partitions for the $L_y=3$ strip.}}
\label{L3partitions}
\end{figure}

To illustrate the method further, we list graphically all the possible
partitions for the strips with $L_y=1$, $L_y=2$, and $L_y=3$ in
Figs. \ref{L1partitions} - \ref{L3partitions}, where white circles are the
original $L_y$ vertices, each black circle corresponds to a specific color
assignment, and the crosses are the vertices in the $q=1$ state. In the
following discussion, we will simply use the names white and black circles and
crosses with the meaning understood. We denote the set of partitions ${\cal
P}_{L_y,d}$ for $1 \le L_y \le 3$ as follows. For simplicity, a single white
circle is not given explicitly in the notation for a partition, where
contiguous vertex numbers are in the same state (color), vertices with overline
are color-assigned, and vertices with underline are in the $q=1$ state. In the
following set of partitions ${\cal P}_{L_y,d}$, individual partitions are
separated by a semicolon. For each partition, vertices that are not in the same
state are separated by a comma. 
\beq 
{\cal P}_{1,0} = \{ I; \underline{1} \} \ , \qquad {\cal P}_{1,1} = 
\{ \bar 1 \} 
\label{L1partitionlist} 
\eeq
\beq 
{\cal P}_{2,0} = \{ I; \underline{1}; \underline{2}; \underline{12}; 12 \}
\ , \qquad {\cal P}_{2,1} = \{ \bar 1; \bar 1, \underline{2}; \bar 2;
\underline{1}, \bar 2; \overline{12} \} \ , \qquad {\cal P}_{2,2} = \{ \bar 1,
\bar 2 \}
\label{L2partitionlist} 
\eeq
\beqs 
{\cal P}_{3,0} & = & \{ I; \underline{1}; \underline{2}; \underline{3};
\underline{12}; \underline{13}; \underline{23}; \underline{123}; 12; 12,
\underline{3}; 13; 13, \underline{2}; 23; \underline{1}, 23; 123 \} \ , \cr\cr
{\cal P}_{3,1} & = & \{ \bar 3; \underline{1}, \bar 3; \underline{2}, \bar 3;
\underline{12}, \bar 3; \bar 2; \underline{1}, \bar 2; \underline{3}, \bar 2;
\underline{13}, \bar 2; \bar 1; \bar 1, \underline{2}; \bar 1, \underline{3};
\bar 1, \underline{23}; 12, \bar 3; \overline{12}; \overline{12},
\underline{3}; \overline{13}; \overline{13}, \underline{2}; \overline{23};
\cr\cr & & \underline{1}, \overline{23}; \bar 1, 23; \overline{123} \} \ ,
\cr\cr {\cal P}_{3,2} & = & \{ \bar 2, \bar 3; \underline{1}, \bar 2, \bar 3;
\bar 1, \bar 3; \bar 1, \bar 3, \underline{2}; \bar 1, \bar 2; \bar 1, \bar 2,
\underline{3}; \overline{12}, \bar 3; \bar 1, \overline{23} \} \ , \qquad {\cal
P}_{3,3} = \{ \bar 1, \bar 2, \bar 3 \}
\label{L3partitionlist} 
\eeqs
\begin{figure}
\unitlength 1mm 
\begin{picture}(50,4)
\put(0,4){\makebox(0,0){{\small $d=0$}}}
\put(10,4){\circle{2}}
\put(20,4){\makebox(0,0){$\times$}}
\put(40,4){\makebox(0,0){{\small $d=1$}}}
\put(50,0){\circle*{2}}
\put(50,4){\circle{2}}
\put(50,0){\line(0,1){4}}
\end{picture}
\caption{\footnotesize{Partitions for the $L_y=1$ strip.}}
\label{L1partitions}
\end{figure}
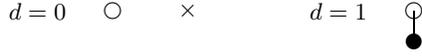
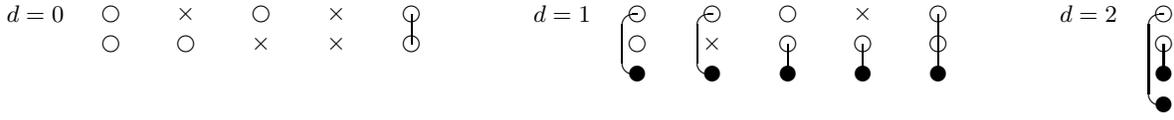
\begin{figure}
\unitlength 1mm 
\begin{picture}(150,12)
\put(0,12){\makebox(0,0){{\small $d=0$}}}
\multiput(10,8)(10,0){2}{\circle{2}}
\multiput(10,12)(20,0){2}{\circle{2}} 
\multiput(30,8)(10,0){2}{\makebox(0,0){$\times$}} 
\multiput(20,12)(20,0){2}{\makebox(0,0){$\times$}} 
\multiput(50,8)(0,4){2}{\circle{2}} 
\put(50,8){\line(0,1){4}}
\put(70,12){\makebox(0,0){{\small $d=1$}}}
\multiput(80,4)(10,0){5}{\circle*{2}}
\put(80,8){\circle{2}}
\multiput(100,8)(10,0){3}{\circle{2}}
\multiput(80,12)(10,0){3}{\circle{2}} 
\put(120,12){\circle{2}}
\multiput(90,8)(20,4){2}{\makebox(0,0){$\times$}} 
\multiput(80,8)(10,0){2}{\oval(4,8)[l]} 
\multiput(100,4)(10,0){2}{\line(0,1){4}}
\put(120,4){\line(0,1){8}}
\put(140,12){\makebox(0,0){{\small $d=2$}}}
\multiput(150,0)(0,4){2}{\circle*{2}}
\multiput(150,8)(0,4){2}{\circle{2}} 
\put(150,4){\line(0,1){4}}
\put(150,6){\oval(4,12)[l]}
\end{picture}
\caption{\footnotesize{Partitions for the $L_y=2$ strip.}}
\label{L2partitions}
\end{figure}

From Eq. (\ref{Tdirectsum}), it follows that the dimension of the total
transfer matrix, i.e., the total number of eigenvalues
$\lambda_{Z,\Lambda,L_y,d,j}$, counting multiplicities, is
\beq
C_{Z,cyc.,L_y} = dim(T_{Z,\Lambda,L_y}) = \sum_{d=0}^{L_y}
dim(T_{Z,\Lambda,L_y,d}) = \sum_{d=0}^{L_y} \tilde c^{(d)} n_{Zh}(L_y,d) = 
q^{L_y} 
\label{dimTL}
\eeq
which is the sum of coefficients, and is independent of the length $m$ of the
strip. We define $N_{Zh,\Lambda,L_y}$ as the total
number of distinct eigenvalues of $T_{Z,\Lambda,L_y}$, i.e.  the sum of the
dimensions of the submatrices $T_{Z,\Lambda,L_y,d}$, modulo the multiplicity
$\tilde c^{(d)}$: 
\beq
N_{Zh,L_y} = \sum_{d=0}^{L_y} n_{Zh}(L_y,d) 
\label{nztot}
\eeq
Here we include an $h$ after the $Z$ in the notation $N_{Zh,\Lambda,L_y}$ to
indicate the presence of a nonzero magnetic field and to avoid confusion with
our earlier notation in Ref. \cite{cf} for the different (smaller) total number
$N_{Z,\Lambda,L_y}$ for the zero-field case.

Using these methods, we have determined the $n_{Zh}(L_y,d)$.  We find that they
are given by the following theorem, whose proof is similar to that for the
zero-field case given in \cite{zt}. 

\bigskip

\begin{theo}
\label{nzlyd}
The dimension of $T_{Z,\Lambda,L_y,d}$, $n_{Zh}(L_y,d)$, 
for $H \ne 0$ and $0 \le d \le L_y$ is determined as follows. One has
\beq
n_{Zh}(L_y,d)=0 \quad {\rm for} \quad d > L_y 
\label{ntup}
\eeq
\beq
n_{Zh}(L_y,L_y)=1
\label{ntcly}
\eeq
and
\beq
n_{Zh}(1,0)=2 
\label{ntc10}
\eeq
All other numbers $n_{Zh}(L_y,d)$ are then determined by the two recursion
relations
\beq
n_{Zh}(L_y+1,0) = 2n_{Zh}(L_y,0) + n_{Zh}(L_y,1)
\label{ntrecursion1}
\eeq
and
\beq
n_{Zh}(L_y+1,d) = n_{Zh}(L_y,d-1) + 3n_{Zh}(L_y,d) + n_{Zh}(L_y,d+1) 
\quad {\rm for} \quad 1 \le d \le L_y+1 
\label{ntrecursion2}
\eeq
\end{theo}
{\it Proof}: Since the maximum number of colors to assign is $L_y$
for a strip with width $L_y$, it follows that $n_{Zh}(L_y,d)=0$ for $d>L_y$ and
$n_{Zh}(L_y,L_y)=1$. It is elementary that $n_{Zh}(1,0)=2$, as shown in
Fig. \ref{L1partitions}. The $d=0$ partitions of a width-$(L_y+1)$ strip can be
obtained by either adding a unconnected white circle or a cross to the bottom
of the $d=0$ partitions of a width-$L_y$ strip, or converting the black circle
of the $d=1$ partitions of a width-$L_y$ strip into a white circle. This gives
Eq. (\ref{ntrecursion1}), which is a special case of the following
discussion. For general $0 \le d \le L_y+1$, the partitions of a
width-$(L_y+1)$ strip can be obtained in one of the following four ways: (a)
for $1 \le d \le L_y+1$, adding a pair of connected circles, one black and one
white, (but not connected to any other vertex) above the highest black circle
of the $d-1$ partitions of a width-$L_y$ strip; (b) for $0 \le d \le L_y$,
adding a unconnected white circle or a cross above the highest black circle of
the $d$ partitions of a width-$L_y$ strip; (c) for $1 \le d \le L_y$, adding a
white circle above the highest black circle of the $d$ partitions of a
width-$L_y$ strip and connecting these two circles; and (d) for $0 \le d \le
L_y-1$, converting the highest black circle of the $d+1$ partitions of a
width-$L_y$ strip into a white circle. Now the lowest white circle of the $d$
partitions of a width-$(L_y+1)$ strip can either connect to a black circle with
or without ((c) or (a)) other connections to other white circles, or it may not
connect to a black circle with or without ((d) or (b)) other connections to
other white circles. The case in which the lowest vertex is a cross is included
in (b). Therefore, (a) to (d) exhaust all the possibilities. The numbers for
these four categories are $n_{Zh}(L_y,d-1)$, $2n_{Zh}(L_y,d)$, $n_{Zh}(L_y,d)$,
and $n_{Zh}(L_y,d+1)$, respectively, so we have
\beqs
n_{Zh}(L_y+1,d) & = & n_{Zh}(L_y,d-1) + 3n_{Zh}(L_y,d) + n_{Zh}(L_y,d+1) 
\quad {\rm for} \ 1 \le d \le L_y-1 \cr\cr
n_{Zh}(L_y+1,L_y) & = & n_{Zh}(L_y,L_y-1) + 3n_{Zh}(L_y,L_y) \cr\cr
n_{Zh}(L_y+1,L_y+1) & = & n_{Zh}(L_y,L_y)
\eeqs
Since $n_{Zh}(L_y,d)=0$ for $d>L_y$, these can be combined into
Eq. (\ref{ntrecursion2}). The relation for $d=0$,
i.e. Eq. (\ref{ntrecursion1}), results from (b) and (d) only. This completes
the proof. \ $\Box$ We remark that these results can also be obtained from 
Eq. (\ref{dimTL}) with $\tilde c^{(d)}$ given in Eq. (\ref{ctd}) by the 
argument in the proof of Theorem 4 in Ref. \cite{cf}. 

\bigskip

A corollary is that
\beq
n_{Zh}(L_y,L_y-1)=3L_y-1 
\label{nzlylyminus1}
\eeq
In Table \ref{ntctablecyc} we list the first few numbers $n_{Zh}(L_y,d)$ and
the total sums $N_{Zh,L_y}$.  In particular, the numbers
$n_{Zh}(L_y,0)$ have been given in Eq. (\ref{nzlyd0}), and the numbers
$n_{Zh}(L_y,1)$ are the binomial transforms of the first differences of the
Catalan numbers. 

Two corollaries of Theorem \ref{nzlyd} are the following. First, 
\beq
N_{Zh,L_y+1}=5N_{Zh,L_y}-2n_{Zh}(L_y,0) 
\label{nttotrecursion}
\eeq
Second, $N_{Zh,L_y}$ can be expressed as 
\beq
N_{Zh,L_y} = \sum_{j=0}^{L_y} {L_y \choose j} {2j \choose j} 
\eeq

\begin{table}
\caption{\footnotesize{Table of numbers $n_{Zh}(L_y,d)$ and their sums, 
$N_{Zh,L_y}$. Blank entries are zero. }}
\begin{center}
\begin{tabular}{|c|c|c|c|c|c|c|c|c|c|c|c|c|}
\hline\hline 
$L_y  \ \backslash \ d$ 
   & 0    & 1    & 2    & 3    & 4    & 5   & 6   & 7  & 8 & 9 &10 & 
$N_{Zh,L_y}$ \\ \hline\hline
1  & 2    & 1    &      &      &      &     &     &    &   &   &   & 3 
\\ \hline
2  & 5    & 5    & 1    &      &      &     &     &    &   &   &   & 11    
\\ \hline
3  & 15   & 21   & 8    & 1    &      &     &     &    &   &   &   & 45   
\\ \hline
4  & 51   & 86   & 46   & 11   & 1    &     &     &    &   &   &   & 195  
\\ \hline
5  & 188  & 355  & 235  & 80   & 14   & 1   &     &    &   &   &   & 873  
\\ \hline
6  & 731  & 1488 & 1140 & 489  & 123  & 17  & 1   &    &   &   &   & 3989 
\\ \hline
7  & 2950 & 6335 & 5397 & 2730 & 875  & 175 & 20  & 1  &   &   &   & 18483 
\\ \hline
8  & 12235& 27352& 25256& 14462& 5530 & 1420& 236 & 23 & 1 &   &   & 86515  
\\ \hline
9  & 51822&119547&117582& 74172& 32472&10026& 2151& 306& 26& 1 &   & 408105 
\\ \hline
10 &223191&528045&546465&372570&181614&64701&16785&3095&385&29 & 1 & 1936881
\\ \hline\hline  
\end{tabular}
\end{center}
\label{ntctablecyc}
\end{table}

The construction of the transfer matrix for each level (i.e., degree) $d$ can be
carried out by methods similar to those for the zero-field transfer matrix
\cite{zt}. Using the basis elements described above
(e.g. Eqs. (\ref{L1partitionlist})-(\ref{L3partitionlist}) for $1 \le L_y \le
3$), we define $J_{L_y,d,i,i+1}$ as the join operator between vertices $i$ and
$i+1$, i.e.
\beq
J_{L_y,d,i,i+1} {\bf v}_{\cal P} = {\bf v}_{{\cal P}i(i+1)} \ , 
\eeq
where both of these vertices $i$ and $i+1$ are in states different from the
$q=1$ state and have not already been assigned different colors. ${\cal
P}i(i+1)$ denotes the partition with $d$ color assignments obtained from $\cal
P$ by connecting vertices $i$ and $i+1$ (regardless of whether they were
already connected or not). For each subspace $d$, we also define $D_{L_y,d,i}$
as the detach operator on vertex $i$ such that
\beq
D_{L_y,d,i} {\bf v}_{\cal P} = \cases{ {\bf v}_{{\cal P}\backslash i} + 
{\bf v}_{{\cal P}\backslash \underline{i}} & if $i$ is a cross or it is 
connected to other vertices \cr
(q-1)( {\bf v}_{{\cal P} i} + {\bf v}_{{\cal P} \underline{i}} ) & 
if $i$ is a white circle without connection } \ , 
\eeq
where the vertex $i$ should not have been assigned color. ${\cal P}\backslash
i$ is the partition obtained from ${\cal P}$ by making $i$ a white circle
without connection, and similarly, ${\cal P}\backslash \underline{i}$ is the
partition obtained from ${\cal P}$ by making $i$ a cross. ${\cal P} i$ is the
same as ${\cal P}$ if $i$ has originally no connection, and ${\cal P}
\underline{i}$ is the partition obtained from ${\cal P}$ by converting $i$ into
a cross. Since the application of these operators does not increase the number
of colors assigned, i.e. $q-1$, the full transfer matrix in
Eq. (\ref{Tdirectsum}) has a triangular block (submatrix) form, and the block
corresponding to a $d$-color assignment has a diagonal block form with $\tilde
c^{(d)}$ blocks.  The transfer matrix $T_{Z,\Lambda,L_y,d}$ for each $d$ is the
product of the transverse and longitudinal parts, $H_{Z,\Lambda,L_y,d}$ and
$V_{Z,\Lambda,L_y,d}$, which can be expressed as
\beqs
H_{Z,sq,L_y,d} & = & H_{Z,tri,L_y,d} = K \prod_{i=1}^{L_y-1}
(I+vJ_{L_y,d,i,i+1}) \cr\cr 
H_{Z,hc,L_y,d,1} & = & K \prod _{i=1}^{[L_y/2]} (I+vJ_{L_y,d,2i-1,2i}) \ , 
\qquad H_{Z,hc,L_y,d,2} = K \prod _{i=1}^{[(L_y-1)/2]} (I+vJ_{L_y,d,2i,2i+1}) 
\cr\cr 
V_{Z,sq,L_y,d} & = & V_{Z,hc,L_y,d} = \prod_{i=1}^{L_y} (vI+D_{L_y,d,i}) \cr\cr
V_{Z,tri,L_y,d} & = & \prod_{i=1}^{L_y-1} [(vI+D_{L_y,d,i})
(I+vJ_{L_y,d,i,i+1})] (vI+D_{L_y,d,L_y}) \ , 
\label{HVmatrix} 
\eeqs
where $[\nu]$ denotes the integral part of $\nu$. Here $K$ is the diagonal
matrix with diagonal element $w^\ell$, where $\ell$ is the number of vertices
in the $q=1$ state for the corresponding basis element.  We have
\beqs
T_{Z,sq,L_y,d} & = & V_{Z,sq,L_y,d} H_{Z,sq,L_y,d} \ , \qquad
T_{Z,tri,L_y,d} = V_{Z,tri,L_y,d} H_{Z,tri,L_y,d} \cr\cr 
T_{Z,hc,L_y,d} & = & (V_{Z,hc,L_y,d} H_{Z,hc,L_y,d,2}) (V_{Z,hc,L_y,d} 
H_{Z,hc,L_y,d,1}) \equiv T_{Z,hc,L_y,d,2} T_{Z,hc,L_y,d,1} 
\label{transfermatrix}
\eeqs

We explain various details for both cyclic and M\"obius strips.  Consider two
adjacent sets of $L_y$ vertices and denote these as $i_1, i_2, ... i_{L_y}$ and
$j_1, j_2, ... j_{L_y}$. For cyclic strips of the square lattice, the
horizontal edges connecting these vertices are $(i_1,j_1)$, $(i_2,j_2)$, ...,
$(i_{L_y},j_{L_y})$. For M\"obius strips, one set of horizontal edges becomes
$(i_1,j_{L_y})$, $(i_2,j_{L_y-1})$, ..., $(i_{L_y},j_1)$. This corresponds to
exchanging the pair of basis elements that switch to each other when the
vertices reverse in order, i.e., the set of basis elements that do not have
self-reflection symmetry with respect to the center of the $L_y$ vertices. For
example, among the partitions for the $L_y=2$ strip in Fig. \ref{L2partitions},
the second partition $\underline 1$ and the third partition $\underline 2$ in
${\cal P}_{2,0}$ must be exchanged under this reflection. Similarly, the pairs
of partitions in ${\cal P}_{2,1}$ are (i) the first partition $\bar 1$ and the
third partition $\bar 2$, (ii) the second partition $\bar 1, \underline 2$ and
the forth partition $\underline 1, \bar 2$. For this specific set of edges of
M\"obius strips, the pairs of columns of $V_{Z,\Lambda,L_y,d}$ that correspond
to these pairs of partitions should be exchanged, and these matrices will be
denoted as $\hat V_{Z,\Lambda,L_y,d}$. Equivalently, the same pairs of
columns of $T_{Z,\Lambda,L_y,d}$ should be exchanged, and these matrices will
be denoted as $\hat T_{Z,\Lambda,L_y,d}=\hat V_{Z,\Lambda,L_y,d}
H_{Z,\Lambda,L_y,d}$ for $\Lambda = sq, tri$.  There are two kinds of M\"obius
strips for the honeycomb lattice.  When $L_y$ is even, the number of vertices
in the horizontal direction is even as for the cyclic strips, i.e., $L_x =
2m$. When $L_y$ is odd, the number of vertices in the horizontal direction is
odd, $L_x = 2m-1$. Therefore, for the honeycomb lattice, we use the definition
\beqs 
\hat T_{Z,hc,L_y,d} & = & \hat V_{Z,hc,L_y,d} H_{Z,hc,L_y,d,1}
\qquad \mbox{for odd} \ L_y \cr\cr 
\hat T_{Z,hc,L_y,d} & = & \hat V_{Z,hc,L_y,d} H_{Z,hc,L_y,d,2} 
V_{Z,hc,L_y,d} H_{Z,hc,L_y,d,1} \qquad \mbox{for even} \ L_y 
\eeqs
As was discussed in \cite{hca} for the crossing-subgraph strips, the square of
each eigenvalue of $\hat T_{Z,hc,L_y,d}$ for odd $L_y$ is an eigenvalue of
the corresponding $T_{Z,hc,L_y,d}$.

We now apply these general methods to determine the structure of the Potts
model partition function in a magnetic field on a lattice strip with M\"obius
boundary conditions.  In the case of zero external magnetic field, we
previously determined the changes of coefficients $c^{(d)}$ when the
longitudinal boundary condition is changed from cyclic to M\"obius
\cite{cf,zt}.  Here we have the same changes of coefficients $\tilde c^{(d)}$
for the square, triangular and honeycomb lattices, as follows:
\beq 
\tilde c^{(0)} \to \tilde c^{(0)} 
\label{cd0tran} 
\eeq
\beq 
\tilde c^{(2k)} \to -\tilde c^{(k-1)} \ , \qquad 1 \le k \le 
\Bigl [ \frac{L_y}{2} \Bigr ] 
\label{cdeventran} 
\eeq
and
\beq 
\tilde c^{(2k+1)} \to \tilde c^{(k+1)} \ ,  \qquad 0 \le k \le 
\Bigl [\frac{L_y-1}{2} \Bigr ] 
\label{cdoddtran} 
\eeq
We thus find the following general structure for the Potts model partition 
function for M\"obius strips: 
\beqs 
Z(\Lambda,L_y \times L_x, Mb,q,v,w) & = &
\tilde c^{(0)}Tr[(T_{Z,\Lambda,L_y,0})^{m-1} 
\hat T_{Z,\Lambda,L_y,0}] \cr\cr 
& & + \sum_{d=0}^{[(L_y-1)/2]} \tilde c^{(d+1)} 
Tr[(T_{Z,\Lambda,L_y,2d+1})^{m-1} \hat T_{Z,\Lambda,L_y,2d+1}] \cr\cr 
& & - \sum_{d=1}^{[L_y/2]} \tilde c^{(d-1)} Tr[(T_{Z,\Lambda,L_y,2d})^{m-1} 
\hat T_{Z,\Lambda,L_y,2d}]
\label{zgsum_transfermb} 
\eeqs
For the square lattice or the honeycomb lattice with $L_y$
even, the eigenvalues of $\hat T_{Z,\Lambda,L_y,d}$ are the same as those of
$T_{Z,\Lambda,L_y,d}$ except for possible changes of signs. The number of
eigenvalues with sign changes is equal to the number of column-exchanges from
$T_{Z,\Lambda,L_y,d}$ to $\hat T_{Z,\Lambda,L_y,d}$. Denote the number of
eigenvalues that are the same for $T_{Z,sq,L_y,d}$ and $\hat T_{Z,sq,L_y,d}$
as $n_{Zh}(sq,L_y,d,+)$, and the number of eigenvalues with different signs as
$n_{Zh}(sq,L_y,d,-)$. It is clear that
\beq 
n_{Zh}(L_y,d) = n_{Zh}(sq,L_y,d,+) + n_{Zh}(sq,L_y,d,-) 
\label{nz}
\eeq
Define
\beq 
\Delta n_{Zh}(sq,L_y,d) \equiv n_{Zh}(sq,L_y,d,+) - n_{Zh}(sq,L_y,d,-) 
\label{deltanz}
\eeq
which gives the number of partitions that have self-reflection symmetry. For
example, among the partitions for the $L_y=2$ strip in Fig. \ref{L2partitions},
the partitions $I$, $\underline{12}$ and $12$ in ${\cal P}_{2,0}$, the fifth
partition $\overline{12}$ in ${\cal P}_{2,1}$, and the partition $\bar 1, \bar
2$ in ${\cal P}_{2,2}$ have self-reflection symmetry. Among the partitions for
the $L_y=3$ strip in Fig. \ref{L3partitions}, those with self-reflection
symmetry include (i) the first partition $I$, the third partition
$\underline{2}$, the sixth partition $\underline{13}$, the eighth partition
$\underline{123}$, the eleven partition $13$, the twelve partition $13,
\underline{2}$ and the fifteenth partition $123$ in ${\cal P}_{3,0}$, (ii) the
fifth partition $\bar 2$, the eighth partition $\bar 2, \underline{13}$, the
sixteenth partition $\overline{13}$, the seventeenth partition $\overline{13},
\underline{2}$, and the twenty first partition $\overline{123}$ in ${\cal
P}_{3,1}$, (iii) the third partition $\bar 1, \bar 3$ and the fourth partition
$\bar 1, \bar 3, \underline{2}$ in ${\cal P}_{3,2}$, and (iv) the partition in
${\cal P}_{3,3}$. We list $\Delta n_{Zh}(sq,L_y,d)$ for $1 \le L_y \le 10$ in
Table \ref{nzpmtable}. The relations between $\Delta n_{Zh}(sq,L_y,d)$ are
\beqs 
\Delta n_{Zh}(sq,2n,0) & = & 2 \Delta n_{Zh}(sq,2n-1,0) - 
\Delta n_{Zh}(sq,2n-2,0) \qquad \mbox{for} \ 1 \le n \cr\cr 
\Delta n_{Zh}(sq,2n+1,0) & = & 2 \Delta n_{Zh}(sq,2n,0) + 
\Delta n_{Zh}(sq,2n,1)  \qquad \mbox{for} \ 0 \le n \cr\cr
\Delta n_{Zh}(sq,2n,2m-1) & = & \Delta n_{Zh}(sq,2n,2m) \cr\cr 
& = & \Delta n_{Zh}(sq,2n-1,2m-1) + \Delta n_{Zh}(sq,2n-1,2m) \cr\cr
& & - \Delta n_{Zh}(sq,2n-2,2m-1) \qquad \mbox{for} \ 1 \le m \le n \cr\cr 
\Delta n_{Zh}(sq,2n+1,m) & = & \Delta n_{Zh}(sq,2n,m-1) + 
\Delta n_{Zh}(sq,2n,m) \cr\cr
& & + \Delta n_{Zh}(sq,2n,m+1) \qquad \mbox{for} \ 1 \le m \le 2n+1 \ , 
\eeqs
where the formal quantity $\Delta n_{Zh}(0,d)=\delta_{d,0}$ is assumed. 
Closed-form expressions are given by
\beqs 
\Delta n_{Zh}(sq,2n,2m) & = & \sum_{j=0}^n {n \choose j} {2j \choose j+m} 
\qquad \mbox{for} \ 0 \le m \le n \cr\cr 
\Delta n_{Zh}(sq,2n+1,m) & = & \cases{ \sum_{j=0}^n {n \choose j} 
{2j \choose j+m/2} \frac{6j+m+4}{2j+m+2} & for even $0 \le m \le 2n$ \cr
\sum_{j=0}^n {n \choose j} {2j \choose j+(m-1)/2} \frac{6j-m+3}{2j+m+1} & 
for odd $1 \le m \le 2n+1$ \cr } 
\eeqs
The total number of these partitions for each $L_y$, denoted as $\Delta
N_{Zh,L_y}$, is
\beq
\Delta N_{Zh,L_y} = \sum_{d=0}^{L_y} \Delta n_{Zh}(sq,L_y,d) = 
\cases{ 5^{L_y/2} & for even $L_y$ \cr
3 \times 5^{(L_y-1)/2} & for odd $L_y$ \cr } 
\eeq

\subsection{Structure of $Z(G,q,v,w)$ for Free Lattice Strips}

Using the methods discussed above, we find that the Potts model partition
function in a magnetic field, on a lattice strips with free boundary
conditions, has the form 
\beqs 
Z(\Lambda, L_y \times L_x, free, q, v, w) & = & u_{L_y}^{\rm T}
H_{Z,\Lambda,L_y,0} (T_{Z,\Lambda,L_y,0})^{L_x-1} s_{L_y} \quad {\rm for} \
\Lambda = sq, tri \cr\cr 
Z(hc, L_y \times L_x, free, q, v, w) & = &
u_{L_y}^{\rm T} H_{Z,hc,L_y,0,1} (T_{Z,hc,L_y,0})^{[(L_x-1)/2]}
(T_{Z,hc,L_y,0,2})^\delta s_{L_y} \ , 
\label{zfree}
\eeqs
where $\delta$ is defined by 
\beq
\delta = \cases{ 1 & for even $L_y$ \cr 0 & for odd $L_y$ } 
\eeq
The element of the vector $u_{L_y}$ for the partition $\cal P$ is given by
$(q-1)^{|{\cal P}|}$, where $|{\cal P}|$ is the number of components for $\cal
P$ that are not in the $q=1$ state. The element of the vector $s_{L_y}$ for the
partition $\cal P$ is equal to unity if there are no connections in $\cal P$,
where certain vertices can be in the $q=1$ state. As an example, $u_2^{\rm T} =
((q-1)^2,q-1,q-1,1,q-1)$ and $s_2^{\rm T} = (1,1,1,1,0)$ for $L_y=2$. For a
strip with free boundary conditions, only the $d=0$ transfer matrix for the
strip with cyclic boundary conditions is needed. For the square lattice or the
honeycomb lattice with $L_y$ even, the size of transfer matrix can be reduced
to $n_{Zh}(sq,L_y,0,+)$ due to reflection symmetry as for the zero-field case
\cite{ts}. We list $n_{Zh}(sq,L_y,d,+)$ and $n_{Zh}(sq,L_y,d,-)$ for $1 \le L_y
\le 10$ in Table \ref{ntctable}. From Eqs. (\ref{nz}) and (\ref{deltanz}),
these are given by
\beqs
n_{Zh}(sq,L_y,d,+) & = & \frac12 \Bigl [ n_{Zh}(L_y,d) + 
\Delta n_{Zh}(sq,L_y,d) \Bigr ] \cr\cr
n_{Zh}(sq,L_y,d,-) & = & \frac12 \Bigl [ n_{Zh}(L_y,d) - 
\Delta n_{Zh}(sq,L_y,d) \Bigr ]
\eeqs

\begin{table}
\caption{\footnotesize{Table of $\Delta n_{Zh}(sq,L_y,d)$ for strips of the
square lattice. Blank entries are zero. The last entry for each value of $L_y$
is the total number of partitions with self-reflection symmetry.}}
\begin{center}
\begin{tabular}{|c|c|c|c|c|c|c|c|c|c|c|c|c|}
\hline\hline 
$L_y \ \backslash \ d$ & 0 & 1 & 2 & 3 & 4 & 5 & 6 & 7 & 8 & 9 & 10 & 
$\Delta N_{Zh,L_y}$ \\ \hline\hline 
1  & 2   & 1   &     &     &     &    &    &    &    &   &   & 3    \\ \hline 
2  & 3   & 1   & 1   &     &     &    &    &    &    &   &   & 5    \\ \hline 
3  & 7   & 5   & 2   & 1   &     &    &    &    &    &   &   & 15   \\ \hline 
4  & 11  & 6   & 6   & 1   & 1   &    &    &    &    &   &   & 25   \\ \hline 
5  & 28  & 23  & 13  & 8   & 2   & 1  &    &    &    &   &   & 75   \\ \hline 
6  & 45  & 30  & 30  & 9   & 9   & 1  & 1  &    &    &   &   & 125  \\ \hline 
7  & 120 & 105 & 69  & 48  & 19  & 11 & 2  & 1  &    &   &   & 375  \\ \hline 
8  & 195 & 144 & 144 & 58  & 58  & 12 & 12 & 1  & 1  &   &   & 625  \\ \hline 
9  & 534 & 483 & 346 & 260 & 128 & 82 & 25 & 14 & 2  & 1 &   & 1875 \\ \hline
10 & 873 & 685 & 685 & 330 & 330 & 95 & 95 & 15 & 15 & 1 & 1 & 3125 \\ \hline\hline 
\end{tabular}
\end{center}
\label{nzpmtable}
\end{table}

\begin{table}
\caption{\footnotesize{Table of numbers $n_{Zh}(sq,L_y,d,\pm)$ for strips of
the square lattice.  For each $L_y$ value, the entries in the first and second
lines are $n_{Zh}(sq,L_y,d,+)$ and $n_{Zh}(sq,L_y,d,-)$, respectively. Blank
entries are zero. The last entry for each value of $L_y$ is the total
$N_{Zh,L_y}$.}}
\begin{center}
\begin{tabular}{|c|c|c|c|c|c|c|c|c|c|c|c|c|}
\hline\hline
\qquad $ (d,+) $            & $0,+$ & $1,+$ & $2,+$ & $3,+$ & $4,+$ & $5,+$ & 
$6,+$ & $7,+$ & $8,+$ & $9,+$ & $10,+$ & $N_{Zh,L_y}$ \\
$L_y \ \backslash \ (d,-) $ & $0,-$ & $1,-$ & $2,-$ & $3,-$ & $4,-$ & $5,-$ & 
$6,-$ & $7,-$ & $8,-$ & $9,-$ & $10,-$ & \\ \hline\hline
1  & 2     & 1     &       &       &       &      &      &     &    &    &   & 3     \\
   &       &       &       &       &       &      &      &     &    &    &   &       \\ \hline
2  & 4     & 3     & 1     &       &       &      &      &     &    &    &   & 11    \\
   & 1     & 2     &       &       &       &      &      &     &    &    &   &       \\ \hline
3  & 11    & 13    & 5     & 1     &       &      &      &     &    &    &   & 45    \\
   & 4     & 8     & 3     &       &       &      &      &     &    &    &   &       \\ \hline
4  & 31    & 46    & 26    & 6     & 1     &      &      &     &    &    &   & 195   \\
   & 20    & 40    & 20    & 5     &       &      &      &     &    &    &   &       \\ \hline
5  & 108   & 189   & 124   & 44    & 8     & 1    &      &     &    &    &   & 873   \\
   & 80    & 166   & 111   & 36    & 6     &      &      &     &    &    &   &       \\ \hline
6  & 388   & 759   & 585   & 249   & 66    & 9    & 1    &     &    &    &   & 3989  \\
   & 343   & 729   & 555   & 240   & 57    & 8    &      &     &    &    &   &       \\ \hline
7  & 1535  & 3220  & 2733  & 1389  & 447   & 93   & 11   & 1   &    &    &   & 18483 \\
   & 1415  & 3115  & 2664  & 1341  & 428   & 82   & 9    &     &    &    &   &       \\ \hline
8  & 6215  & 13748 & 12700 & 7260  & 2794  & 716  & 124  & 12  & 1  &    &   & 86515 \\
   & 6020  & 13604 & 12556 & 7202  & 2736  & 704  & 112  & 11  &    &    &   &       \\ \hline
9  & 26178 & 60015 & 58964 & 37216 & 16300 & 5054 & 1088 & 160 & 14 & 1  &   &408105 \\
   & 25644 & 59532 & 58618 & 36956 & 16172 & 4972 & 1063 & 146 & 12 &    &   &       \\ \hline
10 & 112032& 264365& 273575& 186450& 90972 & 32398& 8440 & 1555& 200& 15 & 1 &1936881\\
   & 111159& 263680& 272890& 186120& 90642 & 32303& 8345 & 1540& 185& 14 &   &       \\ 
\hline\hline
\end{tabular}
\end{center}
\label{ntctable}
\end{table}

As we did for the zero-field case in Ref. \cite{cf}, we define, for the case of
nonzero field, the numbers of $\lambda_{Z,sq,L_y,j}$ for the M\"obius strips of
the square lattice with coefficients $\pm c^{(d)}$ as
$n_{Zh,Mb}(sq,L_y,d,\pm)$. We list $n_{Zh,Mb}(sq,L_y,d,+)$ and
$n_{Zh,Mb}(sq,L_y,d,-)$ for $1 \le L_y \le 10$ in Table \ref{ntctablemb}. With
the Eqs.  (\ref{cd0tran}) to (\ref{cdoddtran}), the relations between
$n_{Zh}(sq,L_y,d,\pm)$ and $n_{Zh,Mb}(sq,L_y,d,\pm)$ are
\beqs 
n_{Zh,Mb}(sq,L_y,0,\pm) & = & n_{Zh}(sq,L_y,0,\pm) + n_{Zh}(sq,L_y,2,\mp) 
\cr\cr
n_{Zh,Mb}(sq,L_y,k,\pm) & = & n_{Zh}(sq,L_y,2k-1,\pm)+n_{Zh}(sq,L_y,2k+2,\mp) 
\cr\cr
& & {\rm for} \quad 1 \le k \le \Bigl [ \frac{L_y+1}{2} \Bigr ]  
\eeqs
The differences for each $d$ are defined as 
\beq
\Delta n_{Zh,Mb}(sq,L_y,d) = n_{Zh,Mb}(sq,L_y,d,+)-n_{Zh,Mb}(sq,L_y,d,-) 
\label{deltan}
\eeq
For the M\"obius strip of the square lattice or the honeycomb lattice with
$L_y$ even, the sign changes of the eigenvalues of $\hat T_{Z,\Lambda,L_y,d}$
can be considered as the sign changes of the coefficients. For these cases, the
sum of coefficients is given by:
\beq
C_{Z,sq,L_y,Mb} \equiv \sum_{j=1}^{N_{Zh,L_y,\lambda}} c_{Z,L_y,Mb,j} = 
\sum_{d=0}^{d_{max}} \Delta n_{Zh,Mb}(sq,L_y,d) \tilde c^{(d)} 
= \cases{ q^{L_y/2} & for even $L_y$ \cr
q^{(L_y+1)/2} & for odd $L_y$ \cr }
\label{ctsummb}
\eeq
where
\beq
d_{max}= \cases{ \frac{L_y}{2} & for even $L_y$ \cr
\frac{(L_y+1)}{2} & for odd $L_y$ \cr }
\label{dmax}
\eeq
That is,
\beq
\Delta n_{Zh,Mb}(sq,2L_y-1,d) = \Delta n_{Zh,Mb}(sq,2L_y,d) = 
n_{Zh}(L_y,d) \quad \mbox{for} \ 0 \le d \le L_y
\eeq

\begin{table}
\caption{\footnotesize{Table of numbers $n_{Zh,Mb}(sq,L_y,d,\pm)$ for M\"obius
strips of the square lattice.  For each $L_y$ value, the entries in the first
and second lines are $n_{Zh,Mb}(sq,L_y,d,+)$ and $n_{Zh,Mb}(sq,L_y,d,-)$,
respectively. Blank entries are zero. The last entry for each value of $L_y$ is
the total $N_{Zh,L_y}$.}}
\begin{center}
\begin{tabular}{|c|c|c|c|c|c|c|c|}
\hline\hline
\qquad $(d,+)$ & $0,+$ & $1,+$ & $2,+$ & $3,+$ & $4,+$ & $5,+$ & $N_{Zh,L_y}$ 
\\
$L_y \ \backslash \ (d,-) $ & $0,-$ & $1,-$ & $2,-$ & $3,-$ & $4,-$ & $5,-$ 
& \\ \hline\hline
1  & 2      & 1      &        &       &      &    & 3       \\
   &        &        &        &       &      &    &         \\ \hline
2  & 4      & 3      &        &       &      &    & 11      \\
   & 2      & 2      &        &       &      &    &         \\ \hline
3  & 14     & 13     & 1      &       &      &    & 45      \\
   & 9      & 8      &        &       &      &    &         \\ \hline
4  & 51     & 46     & 6      &       &      &    & 195     \\
   & 46     & 41     & 5      &       &      &    &         \\ \hline
5  & 219    & 195    & 44     & 1     &      &    & 873     \\
   & 204    & 174    & 36     &       &      &    &         \\ \hline
6  & 943    & 816    & 249    & 9     &      &    & 3989    \\
   & 928    & 795    & 241    & 8     &      &    &         \\ \hline
7  & 4199   & 3648   & 1398   & 93    & 1    &    & 18483   \\
   & 4148   & 3562   & 1352   & 82    &      &    &         \\ \hline
8  & 18771  & 16484  & 7372   & 716   & 12   &    & 86515   \\
   & 18720  & 16398  & 7326   & 705   & 11   &    &         \\ \hline
9  & 84796  & 76187  & 38279  & 5066  & 160  & 1  & 408105  \\
   & 84608  & 75832  & 38044  & 4986  & 146  &    &         \\ \hline
10 & 384922 & 355007 & 194795 & 32583 & 1555 & 15 & 1936881 \\
   & 384734 & 354652 & 194560 & 32503 & 1541 & 14 &         \\ 
\hline\hline
\end{tabular}
\end{center}
\label{ntctablemb}
\end{table}

In previous work we have given zero-field results for the determinants for
various strip graphs $G_s$ (e.g., \cite{s3a})
\beq 
\det T_Z(G_s) = \prod_{j=1}^{N_{Z,G_s,\lambda}}
(\lambda_{Z,G_s,j})^{c_{Z,G_s,j}}
\label{detformz}
\eeq
where $c_{Z,G_s,j}$ is the multiplicity of $\lambda_{Z,G_s,j}$. In the present
context, these can be written, for cyclic strips, as
\beq
det(T_{Z,\Lambda,L_y}) = \prod_{d=0}^{L_y} [det(T_{Z,\Lambda,L_y,d})]^{\tilde
c^{(d)}}
\label{detformzcyc}
\eeq
and we shall extend these results to arbitrary width with a general, nonzero
magnetic field below.

\section{Properties of Transfer Matrices at Special Values of Parameters}

In this section we derive some properties of the transfer matrices
$T_{Z,\Lambda,L_y,d}$ at special values of $q$, $v$, and $w$.  Some related
factorizations were given in \cite{hl}.

\subsection{ $v=0$ }

From (\ref{clusterw}) it follows that for any graph $G$, the Potts model
partition function $Z(G,q,v,w)$ at $v=0$ satisfies
\beq
Z(G,q,0,w)=(q-1+w)^{n(G)} 
\label{zv0}
\eeq
Since this holds for arbitrary values of $q$, in the context of the lattice
strips considered here, it implies
\beq
(T_{Z,\Lambda,L_y,d})_{v=0} = 0 \quad {\rm for} \quad 1 \le d \le L_y
\label{tzlydv0}
\eeq
i.e. these are zero matrices. Secondly, restricting to cyclic strips for
simplicity, and using the basic results $n=L_yL_x=L_y m$ for $\Lambda=sq,tri$
and $n=2L_y m$ for $\Lambda=hc$, Eq. (\ref{zv0}) implies that
\beq
Tr[(T_{Z,\Lambda,cyc.,L_y})^m]_{v=0} 
=\cases{(q-1+w)^{L_y m}  & \ for \ $\Lambda=sq,tri$ \cr
        (q-1+w)^{2L_y m} & \ for \ $\Lambda=hc$  }
\label{TThc1xl} 
\eeq
With our explicit calculations, we find that all of the eigenvalues of the
matrix $T_{Z,\Lambda,L_y,d=0}$ for $v=0$ vanish except for one, which is equal
to $(q-1+w)^{L_y}$ if $\Lambda=sq, tri$ and $(q-1+w)^{2L_y}$ if
$\Lambda=hc$. As will be seen, this is reflected in the property that
$det(T_{Z,\Lambda,L_y,d})$ has a nonzero power of $v$ as a factor for $L_y \ge
2$ for all of the lattice strips considered here. The restriction $L_y \ge 2$
is made because the strips of the triangular and honeycomb lattice are only 
well-defined without degenerating for $L_y \ge 2$. In the case of the square
lattice, for the case $L_y=1$, the transfer matrices $T_{Z,sq,1,0}$ also has
one eigenvalue $q-1+w$ and the other one equal to zero at $v=0$, and
$T_{Z,sq,1,1}$ is the scalar $v$.

\subsection{$v=-1$}

The special case $v=-1$ defines two new types of weighted graph coloring
problems, as we have discussed in \cite{hl}.  We recall that the chromatic
polynomial $P(G,q)$ counts the number of ways of assigning $q$ colors to the
vertices of a graph $G$ such that no adjacent vertices have the same color.
This ``proper $q$-coloring'' of the vertices of $G$ is equivalent to $Z$ for
the zero-temperature, zero-field Potts antiferromagnet, $v=-1$:
$P(G,q)=Z(G,q,-1)$.  Here we have a generalization of this to a weighted proper
$q$-coloring of the vertices of $G$, as described by the polynomial \cite{hl}
\beq
Ph(G,q,w) = Z(G,q,-1,w)
\label{ph}
\eeq
For $H < 0$, i.e., $0 \le w < 1$, this is a weighted graph coloring problem in
which one carries out a proper $q$-coloring of the vertices of $G$ but with a
penalty factor of $w$ for each vertex assigned the color 1.  For $H > 0$, this
is a second type of weighted graph coloring problem, namely a proper vertex $q$
coloring with a weighting that favors one color.  Since this favoring of one
color conflicts with the strict constraint that no two adjacent vertices have
the same color, the range $w > 1$ involves competing interactions and
frustration.  In the limit $w \to \infty$, it is impossible to satisfy the
proper coloring constraint, and this is embodied in the analytic result that
for large positive $w$, $Z(G,q,v,w) \sim (v+1)^{e(G)}w^{n(G)}$, which vanishes
as $v \to -1$.  In this $v=-1$ special case, there are reductions in the ranks
of the transfer matrices $T_{Z,\Lambda,L_y,d}$, i.e., some of the eigenvalues
vanish.  This yields a new set of dimensions of matrix blocks, $n_{Ph}(L_y,d)
\le n_{Zh}(L_y,d)$.  This is a strict inequality, i.e., $n_{Ph}(L_y,d) <
n_{Zh}(L_y,d)$, for all cases except $d=L_y$, where
$n_{Ph}(L_y,L_y)=1=n_{Zh}(L_y,L_y)$.  We have calculated these $n_{Ph}(L_y,d)$
and have obtained a number of interesting properties of the weighted graph
coloring polynomial $Ph(L_y,d)$.  These are beyond the scope of the present
work and hence will be presented elsewhere.

\subsection{$q=0$}

By substituting $q=0$ in (\ref{clusterw}) and noting the factorization
$w^{n(G'_i)}-1 = (w-1)\sum_{\ell=0}^{n(G'_i)-1} w^\ell$, we obtain the result
that $Z(G,0,v,w)$ contains a factor of $(w-1)$.

\subsection{ $q=1$ } 

Evaluating Eq. (\ref{z}) for $q=1$, one sees that the Kronecker delta
functions $\delta_{\sigma_i \sigma_j}=1$ for all pairs of adjacent vertices
$\langle i,j \rangle$; consequently,
\beq
Z(G,1,v,w)=e^{Ke(G)+hn(G)} = (v+1)^{e(G)}w^{n(G)}  
\label{zq1}
\eeq
The coefficients $\tilde c^{(d)}$ evaluated at $q=1$ satisfy \cite{cf}
\beq
\tilde c^{(d)}(q=1) = (-1)^d  
\label{cdq1}
\eeq
Hence, in terms of transfer matrices, we derive the following sum rule for the 
present cyclic lattice strips $G=\Lambda, L_y \times L_x,cyc.$ 
\beq
\sum_{0 \le d \le L_y, \ d \ {\rm even}}  Tr[(T_{Z,\Lambda,L_y,d})^m] - 
\sum_{1 \le d \le L_y, \ d \ {\rm odd}}   Tr[(T_{Z,\Lambda,L_y,d})^m] =
(v+1)^{e(G)}w^{n(G)} \quad {\rm for} \ \ q=1 
\label{zq1tran}
\eeq
Here the number of edges $e(G)$ for each type of cyclic strip is
\beq
e(G) = \cases{ (2L_y-1)m & if $\Lambda=sq$ \cr
               (3L_y-2)m & if $\Lambda=tri$ \cr
               (3L_y-1)m & if $\Lambda=hc$  \cr
               2L_ym     & if $\Lambda=G_D$ }
\label{edges}
\eeq
where $m$ is given in terms of $L_x$ by Eq. (\ref{lxm}).  Since it applies for
arbitrary $m$, the sum rule (\ref{zq1tran}) implies relations between the
eigenvalues of the various transfer matrices $T_{Z,\Lambda,L_y,d}$.  

\subsection{$w=1$}

An interesting question concerns how $Z(G,q,v,w)$ reduces when $H \to 0$ (i.e.,
$w \to 1$).  As will be evident from our explicit calculations, for generic $q$
and $v$, various $\lambda_{Z,L_y,d,j}$'s in a given degree-$d$ subspace become
equal to $\lambda_{Z,L_y,d',j}$'s in a subspace of different degree, $d'$.
This process gives rise to ``transmigration'' of $\lambda_{Z,L_y,d,j}$'s; as
we combine the term(s) $\tilde c^{(d)} [\lambda_{Z,L_y,d',j}]^m$ from the
degree-$d$ subspace(s) with the term $\tilde c^{(d')}[\lambda_{Z,L_y,d',j}]^m$
in the degree-$d'$ subspace, this has the effect of yielding the zero-field
term $c^{(d')}[\lambda_{Z,L_y,d',j}]^m$.  Associated with this transmigration
process, the $n_{Zh}(L_y,d)$ are changed to the $n_Z(L_y,d)$ given in Theorem 4
and Table 3 of \cite{cf} (see also \cite{saleur}). 

\subsection{$w=0$}

The special case $w=0$ is described by the relation (\ref{zw10}), which is
valid for an arbitrary graph $G$.  From (\ref{cluster}), it follows that
$Z(G,q,v)$ contains a factor of $q$. From (\ref{zw10}) it therefore follows
that $Z(G,q,v,0)$ contains a factor of $(q-1)$.

\section{General Results for Cyclic Strips of the Square Lattice}

In this section and the subsequent ones we present general results that we have
obtained for the Potts model in an external magnetic field, valid for
arbitrarily large strip width $L_y$ (as well as arbitrarily great length) for
transfer matrices and their properties.  We begin with strips of the square
lattice.

\subsection{Determinants}

We find
\beq
det(T_{Z,sq,L_y,d}) = (v^{L_y})^{n_{Zh}(L_y,d)} \biggl [ w^{L_y} 
\Bigl ( 1+\frac{q}{v} \Bigr )^{L_y} (1+v)^{L_y-1} \biggr ]^{n_{Zh}(L_y-1,d)} 
\label{detTZsqld}
\eeq
where $n_{Zh}(L_y,d)$ was given in Theorem \ref{nzlyd}. This applies for all
$d$, i.e., $0 \le d \le L_y$ with $n_{Zh}(L_y-1,d)=0$ for $d > L_y-1$.  The
factor of $w$ in Eq. (\ref{detTZsqld}) originates from the diagonal matrix $K$
in Eq. (\ref{HVmatrix}), and the power of $w$ is the sum of the number of
vertices in the $q=1$ state of all the $(L_y,d)$-partitions, which is the same
as the power of $(1+q/v)$.

Next, taking into account that the generalized multiplicity of each
$\lambda_{Z,sq,L_y,d}$ is $\tilde c^{(d)}$, we have, for the total determinant,
\beqs
det(T_{Z,sq,L_y}) &=& \prod_{d=0}^{L_y} [det(T_{Z,sq,L_y,d})]^{\tilde c^{(d)}}
\cr\cr &=& \prod_{d=0}^{L_y} [ v^{L_y}]^{n_{Zh}(L_y,d) \tilde c^{(d)}} \Bigl
[w^{L_y} \Bigl ( 1+\frac{q}{v} \Bigr )^{L_y} (1+v)^{L_y-1} \Bigr
]^{n_{Zh}(L_y-1,d) \tilde c^{(d)}} \cr\cr & = & [v^{L_y}]^{\sum_{d=0}^{L_y}
n_{Zh}(L_y,d) \tilde c^{(d)}} \Bigl [ w^{L_y} \Bigl ( 1+\frac{q}{v} \Bigr )^{L_y}
(1+v)^{L_y-1} \Bigr ]^{\sum_{d=0}^{L_y} n_{Zh}(L_y-1,d) \tilde c^{(d)}}
\label{detTZsqLaux}
\eeqs
Using Eq. (\ref{dimTL}) together with $n_{Zh}(L_y,d) = 0$ for $d > L_y$, so 
that $\sum_{d=0}^{L_y} n_{Zh}(L_y-1,d) \tilde c^{(d)}=
\sum_{d=0}^{L_y-1} n_{Zh}(L_y-1,d) \tilde c^{(d)}$, we have
\beq
det(T_{Z,sq,L_y}) = [v^{L_y}]^{q^{L_y}} \Bigl [ w^{L_y} \Bigl ( 1+\frac{q}{v}
 \Bigr )^{L_y} (1+v)^{L_y-1} \Bigr ]^{q^{L_y-1}} 
\label{detTZsqL}
\eeq
This determinant of the transfer matrix for the cyclic strip of the square
lattice applies for arbitrary width $L_y$. It is the generalization, to $H \ne
0$, of the zero-field result given in \cite{zt} with the extra $w$ factor.

\subsection{Eigenvalue for $d=L_y$ for $\Lambda=sq,tri,hc$}

It was shown earlier that the $\lambda$'s for the zero-field Potts model
partition function are the same for a given lattice strip with cyclic, as
compared with M\"obius, boundary conditions \cite{a,tor}.  Indeed, this had
been observed earlier for the special case of the chromatic polynomial, $v=-1$
\cite{bds}-\cite{pt} (and it was shown that the $\lambda$'s for a
strip with Klein bottle boundary conditions are a subset of the $\lambda$'s for
the same strip with torus boundary conditions \cite{tk,tor}).  From Theorem
\ref{nzlyd} one knows that there is only one $\lambda$ for degree $d=L_y$,
which we denote as $\lambda_{Z,\Lambda,L_y,L_y}$. That is, for this value of
$d$, the transfer matrix reduces to $1 \times 1$, i.e. a scalar. We found that
for a cyclic or M\"obius strip of the square, triangular, or honeycomb lattice
with width $L_y$,
\beq
\lambda_{Z,\Lambda,L_y,L_y}=v^{L_y} \quad {\rm for} \quad \Lambda=sq,tri
\label{lamdlysqtri}
\eeq
\beq
\lambda_{Z,\Lambda,L_y,L_y}=v^{2L_y} \quad {\rm for} \quad \Lambda=hc  
\label{lamdlyhc}
\eeq
which are the same as those for the zero-field case \cite{zt}.

\subsection{Transfer Matrix for $d=L_y-1$, $\Lambda=sq$}

From Eq. (\ref{nzlylyminus1}) it follows that for $\Lambda=sq,tri$ or $hc$, the
number of $\lambda_{Z,\Lambda,L_y,d,j}$, for $d=L_y-1$ is
$n_{Zh}(L_y,L_y-1)=3L_y-1$, i.e. the transfer matrix in this subspace,
$T_{Z,\Lambda,L_y,L_y-1}$, is a (square) $(3L_y-1)$-dimensional matrix.  For
$L_y=1$,
\beq
T_{Z,sq,1,0}= \left( \begin{array}{cc}
    v+q-1 & w \\
    q-1   & w(1+v) \end{array} \right ) 
\label{TTsq10}
\eeq 
For $L_y \ge 2$ we find the following general formula.
\beq
(T_{Z,sq,L_y,L_y-1})_{j,j} = v^{L_y-1}(2v+q-1) \quad {\rm for} \quad j = 1 \ \ {\rm and} \ \ j=2L_y-1
\label{TTsqx1}
\eeq
\beq
(T_{Z,sq,L_y,L_y-1})_{2j-1,2j-1} = v^{L_y-1}(3v+q-1) \quad {\rm for} \quad
L_y \ge 3 \ \ {\rm and} \ \ 2 \le j \le L_y-1
\label{TTsqx2}
\eeq
\beq
(T_{Z,sq,L_y,L_y-1})_{2j,2j} = v^{L_y-1}w(1+v) \quad {\rm for} \quad
1 \le j \le L_y
\label{TTsqx3}
\eeq
\beq
(T_{Z,sq,L_y,L_y-1})_{j+1,j} = v^{L_y-1}(v+q-1) \quad {\rm for} \quad j = 1 \ \ {\rm and} \ \ j=2L_y-1
\label{TTsqx4}
\eeq
\beq
(T_{Z,sq,L_y,L_y-1})_{2j,2j-1} = v^{L_y-1}(2v+q-1) \quad {\rm for} \quad
L_y \ge 3 \ \ {\rm and} \ \ 2 \le j \le L_y-1
\label{TTsqx5}
\eeq
\beq
(T_{Z,sq,L_y,L_y-1})_{2j-1,2j} = v^{L_y-1}w \quad {\rm for} \quad
1 \le j \le L_y
\label{TTsqx6}
\eeq
\beqs
& & (T_{Z,sq,L_y,L_y-1})_{2j-1,2j+1} = (T_{Z,sq,L_y,L_y-1})_{2j,2j+1} \cr\cr
& = & (T_{Z,sq,L_y,L_y-1})_{2j+1,2j-1} = (T_{Z,sq,L_y,L_y-1})_{2j+2,2j-1} = v^{L_y} \quad {\rm for} \quad 1 \le j \le L_y-1
\label{TTsq1a}
\eeqs
\beq
(T_{Z,sq,L_y,L_y-1})_{j,j} = v^{L_y}(1+v) \quad {\rm for} \quad 2L_y+1 \le j \le 3L_y-1
\label{TTsqy}
\eeq
\beqs
& & (T_{Z,sq,L_y,L_y-1})_{2j-1,2L_y+j} = (T_{Z,sq,L_y,L_y-1})_{2j,2L_y+j} \cr\cr
& = & (T_{Z,sq,L_y,L_y-1})_{2j+1,2L_y+j} = (T_{Z,sq,L_y,L_y-1})_{2j+2,2L_y+j} = v^{L_y-1}(1+v) \quad {\rm for} \quad 1 \le j \le L_y-1 \cr & &
\label{TTsqwa}
\eeqs
\beq
(T_{Z,sq,L_y,L_y-1})_{2L_y+j,2j-1} = (T_{Z,sq,L_y,L_y-1})_{2L_y+j,2j+1}= v^{L_y+1} \quad {\rm for} \quad 1 \le j \le L_y-1
\label{TTsqva}
\eeq
with all other elements equal to zero.  Thus, $T_{Z,sq,L_y,L_y-1}$ 
consists of four submatrices:

\begin{enumerate}

\item

an upper left square submatrix with indices $i,j$ in the ranges $1 \le i,j \le 2L_y$ and nonzero elements given by Eqs. (\ref{TTsqx1})-(\ref{TTsq1a}) 

\item 

a lower right square submatrix with indices in the ranges $2L_y+1 \le i,j \le 3L_y-1$ and nonzero elements given by Eqs. (\ref{TTsqy})

\item 

an upper right rectangular submatrix with nonzero elements given by Eq. (\ref{TTsqwa})

\item 

a lower left
rectangular submatrix with nonzero elements given by Eq. (\ref{TTsqva})

\end{enumerate} 

For general $q$, $v$ and $w$, $T_{Z,sq,L_y,L_y-1}$ has rank equal to its
dimension, $3L_y-1$.  We illustrate these general formulas for the cases
$L_y=2$ and $L_y=3$. For this purpose we introduce the abbreviations
\beq
v_j=j+v \ , \quad x_j=jv+q-1 
\label{abbrev1}
\eeq
where $j$ is a positive integer.  We have 
\beq
T_{Z,sq,2,1} = v \left( \begin{array}{ccccc}
x_2 & w    & v   & 0    & v_1 \\
x_1 & wv_1 & v   & 0    & v_1 \\    
v   & 0    & x_2 & w    & v_1 \\
v   & 0    & x_1 & wv_1 & v_1 \\       
v^2 & 0    & v^2 & 0    & vv_1 
\end{array} \right )
\label{TTsq21}
\eeq
and
\beq
T_{Z,sq,3,2} = v^2 \left( \begin{array}{cccccccc}
x_2 & w    &  v  & 0    & 0   & 0    & v_1  & 0   \\
x_1 & wv_1 &  v  & 0    & 0   & 0    & v_1  & 0   \\ 
v   & 0    & x_3 & w    & v   & 0    & v_1  & v_1 \\
v   & 0    & x_2 & wv_1 & v   & 0    & v_1  & v_1 \\ 
0   & 0    & v   & 0    & x_2 & w    & 0    & v_1 \\
0   & 0    & v   & 0    & x_1 & wv_1 & 0    & v_1 \\ 
v^2 & 0    & v^2 & 0    & 0   & 0    & vv_1 & 0   \\
0   & 0    & v^2 & 0    & v^2 & 0    & 0    & vv_1     \end{array} \right ) 
\label{TTsq32}
\eeq

In general, neglecting the $v^{L_y-1}$ factor, the upper left-hand submatrix
has a main $2 \times 2$ block diagonal with end blocks equal to $\left(
\begin{array}{cc} 2v+q-1 & w \\ v+q-1 & w(1+v) \end{array} \right )$ and
interior blocks equal to $\left( \begin{array}{cc} 3v+q-1 & w \\ 2v+q-1 &
w(1+v) \end{array} \right )$.  Adjacent to this main block diagonal are two
block diagonals equal to $\left( \begin{array}{cc} v & 0 \\ v & 0 \end{array}
\right )$, and the rest of the submatrix is comprised of triangular regions
filled with 0's.  The upper right-hand submatrix has a band of two $2 \times 1$
diagonals $\left( \begin{array}{c} 1+v \\ 1+v \end{array} \right )$ together
with triangular regions filled with 0's.  The lower left-hand submatrix has a
band of two $1 \times 2$ diagonals $\left( \begin{array}{cc} v^2 & 0
\end{array} \right )$ together with triangular regions filled with 0's.  And
finally, in the right-hand lower submatrix the entries on the main diagonal are
equal to $v(1+v)$ and the rest of this submatrix is made up of triangular
regions of 0's.  For the lowest values $L_y=1,2$, some of these parts, such as
the triangular regions of zeros, are not present.

Having determined the general form of $T_{Z,sq,L_y,d}$ for $d=L_y-1$, we
find that a pair of its eigenvalues $\lambda_{Z,sq,L_y,L_y-1,j}$ are roots of the following quadratic equation,
\beq
x^2 - v^{L_y-1}(vw+w+v+q-1)x + wv^{2L_y-1}(v+q)=0 
\eeq

As corollaries of our general result for $T_{Z,sq,L_y,L_y-1}$ we calculate the trace and determinant.  
\beq
det(T_{Z,sq,L_y,L_y-1}) = v^{L_y(3L_y-1)} w^{L_y} \Bigl ( 1+\frac{q}{v} \Bigr )^{L_y} (1+v)^{L_y-1} 
\label{detTsqdm1}
\eeq
which is a special case of Eq. (\ref{detTZsqld}), and 
\beq
Tr(T_{Z,sq,L_y,L_y-1}) = v^{L_y-1} \Bigl [ (q-1)L_y + v^2(L_y-1) + v(4L_y-3) + w(1+v)L_y \Bigr ]  
\label{traceTsqdm1}
\eeq

\section{General Results for Cyclic Strips of the Triangular Lattice}

\subsection{Determinants}

We find
\beq
det(T_{Z,tri,L_y,d}) = (v^{L_y})^{n_{Zh}(L_y,d)} \biggl [ w^{L_y} \Bigl (
1+\frac{q}{v} \Bigr )^{L_y} (1+v)^{2(L_y-1)} \biggr ]^{n_{Zh}(L_y-1,d)}
\label{detTZtrild}
\eeq
Comparing $V_{Z,sq,L_y,d}$ and $V_{Z,tri,L_y,d}$ in Eq. (\ref{HVmatrix}), one
sees that a set of $(I+vJ_{L_y,d,i,i+1})$ has been included for the triangular
lattice, so that the power of $(1+v)$ becomes twice of the corresponding power
for the square lattice.

Taking into account that the multiplicity of each $\lambda_{Z,tri,L_y,d,j}$ is
$\tilde c^{(d)}$, it follows that the total determinant is
\beq 
det(T_{Z,tri,L_y}) \equiv \prod_{d=0}^{L_y} [det(T_{Z,tri,L_y,d})]^{\tilde
c^{(d)}} = [v^{L_y}]^{q^{L_y}} \biggl [ w^{L_y} \Bigl ( 1+\frac{q}{v} \Bigr
)^{L_y} (1+v)^{2(L_y-1)} \biggr ]^{q^{L_y-1}}
\label{detTZtri}
\eeq

\subsection{Transfer Matrix for $d=L_y-1$, $\Lambda=tri$}

From Eq. (\ref{nzlylyminus1}), we know that the dimension of this transfer
matrix is $3L_y-1$ as before. The first nontrivial case is $L_y=2$. For $L_y
\ge 2$ we find the following general formula.
\beq
(T_{Z,tri,L_y,L_y-1})_{1,1} = v^{L_y-1}(v^2+4v+q-1)
\label{TTtri_11}
\eeq
\beq
(T_{Z,tri,L_y,L_y-1})_{2j-1,2j-1} = v^{L_y-1}(v^2+5v+q-1) \quad {\rm for} \quad 2 \le j \le L_y-1
\label{TTtri_lud}
\eeq
\beq
(T_{Z,tri,L_y,L_y-1})_{2L_y-1,2L_y-1} = v^{L_y-1}(2v+q-1)
\label{TTtri_x2}
\eeq
\beq
(T_{Z,tri,L_y,L_y-1})_{2j,2j} = v^{L_y-1}w(1+v) \quad {\rm for} \quad 1 \le j \le L_y
\label{TTtri_lud2}
\eeq
\beq
(T_{Z,tri,L_y,L_y-1})_{2,1} = v^{L_y-1}(v^2+3v+q-1)
\label{TTtri_21}
\eeq
\beq
(T_{Z,tri,L_y,L_y-1})_{2j,2j-1} = v^{L_y-1}(v^2+4v+q-1) \quad {\rm for} \quad 2 \le j \le L_y-1
\label{TTtri_lud3}
\eeq
\beq
(T_{Z,tri,L_y,L_y-1})_{2L_y,2L_y-1} = v^{L_y-1}(v+q-1)
\label{TTtri_x1}
\eeq
\beq
(T_{Z,tri,L_y,L_y-1})_{2j+1,2j-1} = (T_{Z,tri,L_y,L_y-1})_{2j+2,2j-1} = v^{L_y} \quad {\rm for} \quad 1 \le j \le L_y-1
\label{TTtri_1}
\eeq
\beq
(T_{Z,tri,L_y,L_y-1})_{j,k} = v^{L_y-1}w \quad {\rm for} \quad k=2, 4, ..., 2L_y \quad {\rm and} \quad 1 \le j \le k-1 
\label{TTtri_lu}
\eeq
\beq
(T_{Z,tri,L_y,L_y-1})_{j,k} = v^{L_y-1}(2v^2+6v+q-1) \quad {\rm for} \quad k=3, 5, ..., 2L_y-3 \quad {\rm and} \quad 1 \le j \le k-1
\label{TTtri_cu}
\eeq
\beq
(T_{Z,tri,L_y,L_y-1})_{j,2L_y-1} = v^{L_y-1}(v^2+3v+q-1) \quad {\rm for} \quad 1 \le j \le 2L_y-2
\label{TTtri_cu2}
\eeq
\beq
(T_{Z,tri,L_y,L_y-1})_{2j+1,2L_y+j} = (T_{Z,tri,L_y,L_y-1})_{2j+2,2L_y+j} = v^{L_y-1}(1+v) \quad {\rm for} \quad 1 \le j \le L_y-1
\label{TTtri_rd}
\eeq
\beq
(T_{Z,tri,L_y,L_y-1})_{j,2L_y+k} = v^{L_y-1}(v+1)(v+2) \quad {\rm for} \quad 1 \le k \le L_y-1 \quad {\rm and} \quad 1 \le j \le 2k
\label{TTtri_ru}
\eeq
\beq
(T_{Z,tri,L_y,L_y-1})_{2L_y+j,2j-1} = v^{L_y+1}(3+v) \quad {\rm for} \quad 1 \le j \le L_y-1 
\label{TTtri_ld}
\eeq
\beq
(T_{Z,tri,L_y,L_y-1})_{2L_y+j,2k-1} = v^{L_y}(2v^2+6v+q-1) \quad {\rm for} \quad 2 \le k \le L_y-1 \quad {\rm and}
\quad 1 \le j \le k-1
\label{TTtri_ll}
\eeq
\beq
(T_{Z,tri,L_y,L_y-1})_{2L_y+j,2L_y-1} = v^{L_y}(v^2+3v+q-1) \quad {\rm for} \quad 1 \le j \le L_y-1
\label{TTtri_cl}
\eeq
\beq
(T_{Z,tri,L_y,L_y-1})_{2L_y+j,k} = v^{L_y}w \quad {\rm for} \quad k=4, 6, ..., 2L_y \quad {\rm and}
\quad 1 \le j \le k/2-1
\label{TTtri_ll2}
\eeq
\beq
(T_{Z,tri,L_y,L_y-1})_{2L_y+j,2L_y+k} = v^{L_y}(v+1)(v+2) \quad {\rm for} \quad 1 \le j \le k \le L_y-1
\label{TTtri_rl}
\eeq
with all other elements equal to zero.  Thus, $T_{Z,tri,L_y,L_y-1}$ can again
be usefully viewed as consisting of various submatrices.

For general $q$, $v$ and $w$, $T_{Z,tri,L_y,L_y-1}$ has rank equal to its
dimension, $3L_y-1$. We illustrate these general formulas with some explicit
examples for $L_y=2$ and $L_y=3$.  For compactness of notation, we use the
abbreviations
\beq
y_j=v^2+jv+q-1 \ , \quad z_j=2v^2+jv+q-1 
\label{abbrev2}
\eeq
where $j$ is a positive integer.  Then 
\beq
T_{Z,tri,2,1} = v \left( \begin{array}{ccccc}
y_4    & w    & y_3  & w    & v_1v_2 \\
y_3    & wv_1 & y_3  & w    & v_1v_2 \\
v      & 0    & x_2  & w    & v_1    \\
v      & 0    & x_1  & wv_1 & v_1    \\ 
v^2v_3 & 0    & vy_3 & wv   & vv_1v_2    
\end{array} \right )
\label{TTtri21}
\eeq
\beq
T_{Z,tri,3,2} = v^2 \left( \begin{array}{cccccccc}
y_4    & w    & z_6    & w    & y_3  & w    & v_1v_2  & v_1v_2  \\
y_3    & wv_1 & z_6    & w    & y_3  & w    & v_1v_2  & v_1v_2  \\
v      & 0    & y_5    & w    & y_3  & w    & v_1     & v_1v_2  \\
v      & 0    & y_4    & wv_1 & y_3  & w    & v_1     & v_1v_2  \\
0      & 0    & v      & 0    & x_2  & w    & 0       & v_1     \\
0      & 0    & v      & 0    & x_1  & wv_1 & 0       & v_1     \\
v^2v_3 & 0    & vz_6   & wv   & vy_3 & wv   & vv_1v_2 & vv_1v_2 \\
0      & 0    & v^2v_3 & 0    & vy_3 & wv   & 0       & vv_1v_2    
\end{array} \right )
\label{TTtri32}
\eeq

As corollaries of our general result for $T_{Z,tri,L_y,L_y-1}$ we calculate the
trace and determinant.
\beq
det(T_{Z,tri,L_y,L_y-1}) = v^{L_y(3L_y-1)} w^{L_y} \Bigl ( 1+\frac{q}{v} \Bigr
)^{L_y} (1+v)^{2(L_y-1)}
\label{detTtridm1}
\eeq
which is a special case of Eq. (\ref{detTZtrild}), and 
\beq
Tr(T_{Z,tri,L_y,L_y-1}) = v^{L_y-1} \Bigl [ (q-1)L_y + v^3(L_y-1) + 4v^2(L_y-1)
+ v(7L_y-6) + w(1+v)L_y \Bigr ]
\label{traceTtridm1}
\eeq

\section{General Results for Cyclic Strips of the Honeycomb Lattice}

\subsection{Determinants}

We find
\beq 
det(T_{Z,hc,L_y,d}) = (v^{2L_y})^{n_{Zh}(L_y,d)} \biggl [ w^{2L_y} \Bigl (
1+\frac{q}{v} \Bigr )^{2L_y} (1+v)^{L_y-1} \biggr ]^{n_{Zh}(L_y-1,d)}  
\label{detTZhcld} 
\eeq
This can be understood as follows: by an argument similar to that given before,
the power of $(1+v)$ is the same as for the square lattice case. Comparing
$T_{Z,sq,L_y,d}$ and $T_{Z,hc,L_y,d}$ in Eq. (\ref{transfermatrix}), one sees
that $V_{Z,hc,L_y,d} = V_{Z,sq,L_y,d}$ has been multiplied twice for the
honeycomb lattice, so that the powers of $v$ and $(1+q/v)$ become twice of the
corresponding powers for the square lattice. In Eq. (\ref{HVmatrix}), both
$H_{Z,hc,L_y,d,1}$ and $H_{Z,hc,L_y,d,2}$ include the matrix $K$ so that the
powers of $w$ also become twice of the corresponding powers for the square
lattice.

Taking into account that the multiplicity of each $\lambda_{Z,hc,L_y,d,j}$ is
$\tilde c^{(d)}$, it follows that the total determinant for the $hc$ lattice is
\beq 
det(T_{Z,hc,L_y}) \equiv \prod_{d=0}^{L_y} [det(T_{Z,hc,L_y,d})]^{\tilde
c^{(d)}} = (v^{2L_y})^{q^{L_y}} \biggl [ w^{2L_y} \Bigl ( 1+\frac{q}{v} \Bigr
)^{2L_y} (1+v)^{L_y-1} \biggr ]^{q^{L_y-1}}
\label{detTZhc} 
\eeq

Summarizing the connections between the determinants of the transfer matrices
for the three lattice strips, $det(T_{Z,tri,L_y,d})$ is related to
$det(T_{Z,sq,L_y,d})$ by the replacement of $(1+v)$ by $(1+v)^2$ and
$det(T_{Z,hc,L_y,d})$ is related to $det(T_{Z,sq,L_y,d})$ by the replacements
of the respective factors $w$ by $w^2$, $v$ by $v^2$ and $(1+\frac{q}{v})$ by
$(1 + \frac{q}{v})^2$. This, together with the fact that $n_{Zh}(L_y,d)$ is the
same for all of these three lattices means that the total determinants
$det(T_{Z,tri,L_y})$ and $det(T_{Z,hc,L_y})$ are related to $det(T_{Z,sq,L_y})$
by the same respective replacements.

\subsection{Transfer Matrix for $d=L_y-1$, $\Lambda=hc$}

From Eq. (\ref{nzlylyminus1}), we know that the dimension of this transfer
matrix is again $3L_y-1$. The first nontrivial case is $L_y=2$. Recall in
Eq. (\ref{transfermatrix}) the transfer matrix for the honeycomb lattice,
$T_{Z,hc,L_y,L_y-1}$, is the product of $T_{Z,hc,L_y,L_y-1,1}$ and
$T_{Z,hc,L_y,L_y-1,2}$. For $L_y \ge 2$ we find the following general formula.
\beq 
(T_{Z,hc,L_y,L_y-1,1})_{2j-1,2j-1} = v^{L_y-1}(2v+q-1) \quad {\rm for} \quad 1
\le j \le L_y-1
\label{TThc1x1} 
\eeq
\beq 
(T_{Z,hc,L_y,L_y-1,1})_{2L_y-1,2L_y-1} = 
\cases{ v^{L_y-1}(2v+q-1) & for $L_y$ even \cr
        v^{L_y-1}(v+q-1)  & for $L_y$ odd  }
\label{TThc1xlb} 
\eeq
\beq 
(T_{Z,hc,L_y,L_y-1,1})_{2j,2j} = v^{L_y-1}w(1+v) \quad {\rm for} \quad 1 \le j
\le L_y
\label{TThc1x2} 
\eeq
\beq 
(T_{Z,hc,L_y,L_y-1,1})_{2j,2j-1} = v^{L_y-1}(v+q-1) \quad {\rm for} \quad 1 \le
j \le L_y-1
\label{TThc1x3} 
\eeq
\beq 
(T_{Z,hc,L_y,L_y-1,1})_{2L_y,2L_y-1} = 
\cases{ v^{L_y-1}(v+q-1) & for $L_y$ even \cr
        v^{L_y-1}(q-1)   & for $L_y$ odd  }
\label{TThc1x3b} 
\eeq
\beq 
(T_{Z,hc,L_y,L_y-1,1})_{2j-1,2j} = v^{L_y-1}w \quad {\rm for} \quad 1 \le j \le
L_y
\label{TThc1x4} 
\eeq
\beqs 
& & (T_{Z,hc,L_y,L_y-1,1})_{4j-1,4j-3} = (T_{Z,hc,L_y,L_y-1,1})_{4j,4j-3} \cr &
= & (T_{Z,hc,L_y,L_y-1,1})_{4j-3,4j-1} = (T_{Z,hc,L_y,L_y-1,1})_{4j-2,4j-1}=
v^{L_y} \quad {\rm for} \quad 1 \le j \le [L_y/2]
\label{TThc1_1} 
\eeqs
\beq 
(T_{Z,hc,L_y,L_y-1,1})_{2L_y+2j-1,2L_y+2j-1} = v^{L_y}(1+v) \quad {\rm for}
\quad 1 \le j \le [L_y/2]
\label{TThc1ya} 
\eeq
\beq 
(T_{Z,hc,L_y,L_y-1,1})_{2L_y+2j,2L_y+2j} = v^{L_y} \quad {\rm for} \quad 1 \le
j \le [(L_y-1)/2]
\label{TThc1yb} 
\eeq
\beqs
& & (T_{Z,hc,L_y,L_y-1,1})_{4j-3,2L_y+2j-1} =
(T_{Z,hc,L_y,L_y-1,1})_{4j-2,2L_y+2j-1} \cr & = &
(T_{Z,hc,L_y,L_y-1,1})_{4j-1,2L_y+2j-1} = (T_{Z,hc,L_y,L_y-1,1})_{4j,2L_y+2j-1}
\cr & = & v^{L_y-1}(v+1) \quad {\rm for} \quad 1 \le j \le [L_y/2]
\label{TThc1wa} 
\eeqs
\beqs 
& & (T_{Z,hc,L_y,L_y-1,1})_{4j-1,2L_y+2j} = (T_{Z,hc,L_y,L_y-1,1})_{4j,2L_y+2j}
\cr & = & (T_{Z,hc,L_y,L_y-1,1})_{4j+1,2L_y+2j} =
(T_{Z,hc,L_y,L_y-1,1})_{4j+2,2L_y+2j} \cr & = & v^{L_y-1} \quad {\rm for} \quad
1 \le j \le [(L_y-1)/2]
\label{TThc1wb} 
\eeqs
\beq
(T_{Z,hc,L_y,L_y-1,1})_{2L_y+2j-1,4j-3} =
(T_{Z,hc,L_y,L_y-1,1})_{2L_y+2j-1,4j-1} = v^{L_y+1} \quad {\rm for} \quad 1 \le
j \le [L_y/2]
\label{TThc1v} 
\eeq
\beq 
(T_{Z,hc,L_y,L_y-1,2})_{1,1} = v^{L_y-1}(v+q-1) 
\label{TThc2x} 
\eeq
\beq 
(T_{Z,hc,L_y,L_y-1,2})_{2j-1,2j-1} = v^{L_y-1}(2v+q-1) \quad {\rm for} \quad 2
\le j \le L_y-1
\label{TThc2x1} 
\eeq
\beq 
(T_{Z,hc,L_y,L_y-1,2})_{2L_y-1,2L_y-1} = 
\cases{ v^{L_y-1}(2v+q-1) & for $L_y$ odd \cr
        v^{L_y-1}(v+q-1)  & for $L_y$ even  }
\label{TThc2x2} 
\eeq
\beq 
(T_{Z,hc,L_y,L_y-1,2})_{2j,2j} = v^{L_y-1}w(v+1) \quad {\rm for} \quad 1 \le j
\le L_y
\label{TThc2x3} 
\eeq
\beq 
(T_{Z,hc,L_y,L_y-1,2})_{2,1} = v^{L_y-1}(q-1) 
\label{TThc2x4} 
\eeq
\beq 
(T_{Z,hc,L_y,L_y-1,2})_{2j,2j-1} = v^{L_y-1}(v+q-1) \quad {\rm for} \quad 
2 \le j \le L_y-1 
\label{TThc2x5} 
\eeq
\beq 
(T_{Z,hc,L_y,L_y-1,2})_{2L_y,2L_y-1} = 
\cases{ v^{L_y-1}(v+q-1) & for $L_y$ odd \cr
        v^{L_y-1}(q-1)   & for $L_y$ even  }
\label{TThc2x6} 
\eeq
\beq 
(T_{Z,hc,L_y,L_y-1,2})_{2j-1,2j} = v^{L_y-1}w \quad {\rm for} \quad 1 \le j 
\le L_y
\label{TThc2x7} 
\eeq
\beqs 
& & (T_{Z,hc,L_y,L_y-1,2})_{4j+1,4j-1} = (T_{Z,hc,L_y,L_y-1,2})_{4j+2,4j-1} \cr
& = & (T_{Z,hc,L_y,L_y-1,2})_{4j-1,4j+1} = (T_{Z,hc,L_y,L_y-1,2})_{4j,4j+1} \cr
& = & v^{L_y} \quad {\rm for} \quad 1 \le j \le [(L_y-1)/2] 
\label{TThc2_1} 
\eeqs
\beq 
(T_{Z,hc,L_y,L_y-1,2})_{2L_y+2j-1,2L_y+2j-1} = v^{L_y} \quad {\rm for} \quad 1 \le j \le [L_y/2] 
\label{TThc2ya} 
\eeq
\beq 
(T_{Z,hc,L_y,L_y-1,2})_{2L_y+2j,2L_y+2j} = v^{L_y}(1+v) \quad {\rm for} \quad 1 \le j \le [(L_y-1)/2] 
\label{TThc2yb} 
\eeq
\beqs
& & (T_{Z,hc,L_y,L_y-1,2})_{4j-3,2L_y+2j-1} =
(T_{Z,hc,L_y,L_y-1,2})_{4j-2,2L_y+2j-1} \cr & = &
(T_{Z,hc,L_y,L_y-1,2})_{4j-1,2L_y+2j-1} = (T_{Z,hc,L_y,L_y-1,2})_{4j,2L_y+2j-1}
\cr & = & v^{L_y-1} \quad {\rm for} \quad 1 \le j \le [L_y/2]
\label{TThc2wa} 
\eeqs
\beqs 
& & (T_{Z,hc,L_y,L_y-1,2})_{4j-1,2L_y+2j} = (T_{Z,hc,L_y,L_y-1,2})_{4j,2L_y+2j}
\cr & = & (T_{Z,hc,L_y,L_y-1,2})_{4j+1,2L_y+2j} =
(T_{Z,hc,L_y,L_y-1,2})_{4j+2,2L_y+2j} \cr & = & v^{L_y-1}(1+v) \quad {\rm for}
\quad 1 \le j \le [(L_y-1)/2]
\label{TThc2wb} 
\eeqs
\beq 
(T_{Z,hc,L_y,L_y-1,2})_{2L_y+2j,4j-1} = (T_{Z,hc,L_y,L_y-1,1})_{2L_y+2j,4j+1} =
v^{L_y+1} \quad {\rm for} \quad 1 \le j \le [(L_y-1)/2]
\label{TThc2v}
\eeq
with all other elements equal to zero.

For general $q$, $v$ and $w$, $T_{Z,hc,L_y,L_y-1}$ has rank equal to its
dimension, $3L_y-1$.  We illustrate these general formulas for the cases
$L_y=2$ and $L_y=3$ (with the same abbreviations as before): 
\beq
T_{Z,hc,2,1,1} = T_{Z,sq,2,1} \ , \qquad
T_{Z,hc,2,1,2} = v \left( \begin{array}{ccccc}
x_1 & w    & 0   & 0    & 1 \\
q-1 & wv_1 & 0   & 0    & 1 \\    
0   & 0    & x_1 & w    & 1 \\
0   & 0    & q-1 & wv_1 & 1 \\       
0   & 0    & 0   & 0    & v 
\end{array} \right )
\label{TThc21parts} 
\eeq
\beqs
T_{Z,hc,3,2,1} & = & v^2 \left( \begin{array}{cccccccc}
x_2 & w    & v   & 0    & 0   & 0    & v_1  & 0 \\
x_1 & wv_1 & v   & 0    & 0   & 0    & v_1  & 0 \\
v   & 0    & x_2 & w    & 0   & 0    & v_1  & 1 \\
v   & 0    & x_1 & wv_1 & 0   & 0    & v_1  & 1 \\
0   & 0    & 0   & 0    & x_1 & w    & 0    & 1 \\
0   & 0    & 0   & 0    & q-1 & wv_1 & 0    & 1 \\
v^2 & 0    & v^2 & 0    & 0   & 0    & vv_1 & 0 \\
0   & 0    & 0   & 0    & 0   & 0    & 0    & v \\
\end{array} \right ) 
\cr\cr
T_{Z,hc,3,2,2} & = & v^2 \left( \begin{array}{cccccccc}
x_1 & w    & 0   & 0    & 0   & 0    & 1 & 0    \\
q-1 & wv_1 & 0   & 0    & 0   & 0    & 1 & 0    \\
0   & 0    & x_2 & w    & v   & 0    & 1 & v_1  \\
0   & 0    & x_1 & wv_1 & v   & 0    & 1 & v_1  \\
0   & 0    & v   & 0    & x_2 & w    & 0 & v_1  \\
0   & 0    & v   & 0    & x_1 & wv_1 & 0 & v_1  \\
0   & 0    & 0   & 0    & 0   & 0    & v & 0    \\
0   & 0    & v^2 & 0    & v^2 & 0    & 0 & vv_1 \\
\end{array} \right )
\label{TThc32parts} 
\eeqs

The $T_{Z,hc,L_y,L_y-1}$ are obtained via Eq. (\ref{transfermatrix})
from these auxiliary matrices.  For example, 
%
\footnotesize
\beqs
\lefteqn{T_{Z,hc,2,1}} \cr
& = & v^2 \left( \begin{array}{ccccc}
x_1(3v+w)+(q-1)^2 & w(x_1+wv_1)   & v(x_2+w)          & 0             & v_1(x_2+w)    \\
x_1(x_1+wv_1)     & w(wv_1^2+q-1) & v(x_1+wv_1)       & 0             & v_1(x_1+wv_1) \\
v(x_2+w)          & 0             & x_1(3v+w)+(q-1)^2 & w(x_1+wv_1)   & v_1(x_2+w)    \\
v(x_1+wv_1)       & 0             & x_1(x_1+wv_1)     & w(wv_1^2+q-1) & v_1(x_1+wv_1) \\
v^3               & 0             & v^3               & 0             & v^2v_1 
\end{array} \right ) \cr & &
\label{TZhc21}
\eeqs
\normalsize

We find that in general, for degree $d=L_y-1$, a pair of its eigenvalues
$\lambda_{Z,hc,L_y,L_y-1,j}$ are roots of the following quadratic equation,
\beq
x^2 - v^{2L_y-2} \Bigl ( w^2(v+1)^2+2(q-1)(v+w)+(q-1)^2+v^2 \Bigr ) x +
w^2v^{4L_y-2}(v+q)^2 = 0 \eeq
The expressions for the other $3L_y-3$ eigenvalues are, in general, more
complicated.

\section{General Results for Cyclic Self-Dual Square-Lattice Strips}

In this section we consider the Potts model for families of self-dual strip
graphs of the square lattice with fixed width $L_y$ and arbitrarily great
length $L_x$, having periodic longitudinal boundary conditions, such that all
vertices on one side of the strip, which we take to be the upper side, are
joined by edges to a single external vertex.  A strip graph of this type will
be denoted generically as $G_D$ and, in more detail, as $G_D(L_y\times
L_x)$. The family of $G_D$ graphs is planar and self-dual.  We recall that for
a planar graph $G_{pl}$, one defines the (planar) dual graph $G_{pl}^*$ as the
graph obtained by replacing each vertex (face) of $G_{pl}$ by a face (vertex)
of $G_{pl}^*$ and connecting the vertices of the resultant $G_{pl}^*$ by edges
when the corresponding faces of $G_{pl}$ have a common edge.  The graph
$G_{pl}$ is self-dual if and only if $G_{pl}= G_{pl}^*$. For zero-field, it is
known that
\beq
Z(G_{pl},q,v,w=1) = v^{e(G_{pl})}q^{-c(G_{pl})} Z(G_{pl}^*,q,\frac{q}{v},w=1) 
\label{zdual}
\eeq
for a planar graph $G_{pl}$.  In general, the graph $G_D(L_y \times L_x)$ has
$n \equiv |V|= L_xL_y+1$ vertices, equal to the number of faces, $f$.  One
motivation for considering the $G_D$ strip graphs is that they exhibit, for any
$L_y$, the self-duality property of the infinite square lattice so that the
zero-field partition function is invariant under $v \to q/v$ by (\ref{zdual}),
aside from a prefactor.

In Ref. \cite{dg} we gave the general form for the zero-field Potts model
partition function $Z(G_D,L_y \times L_x,q,v,w=1)$.  In our current notation
with $m$ given in terms of $L_x$ by Eq. (\ref{lxm}), this is
\beq
Z(G_D,L_y \times L_x,q,v,1) = \sum_{d=1}^{L_y+1} \kappa^{(d)}
Tr[(T_{Z,G_D,L_y,d})^m]
\label{zsdgtranw1}
\eeq
where
\beq
\kappa^{(d)} = \sqrt{q} \ U_{2d-1} \Big ( \frac{\sqrt{q}}{2} \Big ) 
= \sum_{j=0}^{d-1} (-1)^j { 2d-1-j \choose j} q^{d-j} \ . 
\label{kappad}
\eeq

To construct the transfer matrix $T_{Z,G_D,L_y,d}$ for each $d$, we 
begin with partitions with $L_y+1$ vertices, where the single external vertex
is considered as the $(L_y+1)$-th vertex, so that the size of the matrix is
$n_{Zh}(L_y+1,d)$ with $0 \le d \le L_y+1$. When the field is non-zero, the
transverse and longitudinal parts, $H_{Z,G_D,L_y,d}$ and $V_{Z,G_D,L_y,d}$, of
the transfer matrix $T_{Z,G_D,L_y,d}$ can be expressed as
\beqs
H_{Z,G_D,L_y,d} & = & K \prod_{i=1}^{L_y} (I+vJ_{L_y,d,i,i+1}) \ , \qquad \bar H_{Z,G_D,L_y,d} = \bar K \prod _{i=1}^{L_y} (I+vJ_{L_y,d,i,i+1}) \cr\cr 
V_{Z,G_D,L_y,d} & = & \prod_{i=1}^{L_y} (vI+D_{L_y,d,i}) \cr\cr
T_{Z,G_D,L_y,d} & = & V_{Z,G_D,L_y,d} H_{Z,G_D,L_y,d} \ , \qquad
\bar T_{Z,G_D,L_y,d} = V_{Z,G_D,L_y,d} \bar H_{Z,G_D,L_y,d}
\label{HVmatrixGD} 
\eeqs
where $K$ is again the diagonal matrix with diagonal element $w^\ell$ where
$\ell$ is the number of vertices in the $q=1$ state for the corresponding
basis, and $\bar K$ is the diagonal matrix with diagonal element $w^{\bar
\ell}$ where $\bar \ell$ is the number of vertices, excluding the $(L_y+1)$-th
vertex, in the $q=1$ state. Notice that $V_{Z,G_D,L_y,d}$ does not include the
factor $(vI+D_{L_y,d,L_y+1})$ for the single external vertex. We have
\beq
Z(G_D, L_y \times L_x,q,v,w) = \sum_{d=0}^{L_y+1} \tilde c^{(d)}
Tr[T_{Z,G_D,L_y,d} (\bar T_{Z,G_D,L_y,d})^{m-1}]
\label{zsdgtran}
\eeq
Here only one $T_{Z,G_D,L_y,d}$ is needed as the single external vertex in the
$q=1$ state should only be considered once. Compare matrices $T_{Z,G_D,L_y,d}$
and $\bar T_{Z,G_D,L_y,d}$, certain columns differ by a factor of $w$,
corresponding to the partitions with the $(L_y+1)$-th vertex in the $q=1$
state. It is clear that the number of these columns is $n_{Zh}(L_y,d)$,
i.e. the cross case in category (b) in the proof of Theorem \ref{nzlyd}. It
follows that there are $n_{Zh}(L_y,d)$ eigenvalues of $T_{Z,G_D,L_y,d}$ equal
to the corresponding eigenvalues of $\bar T_{Z,G_D,L_y,d}$ multiplied by $w$,
and the rest of the eigenvalues are the same. Let us denote the eigenvalues
that are common to $T_{Z,G_D,L_y,d}$ and $\bar T_{Z,G_D,L_y,d}$ as usual as
$\lambda_{Z,G_D,L_y,d,j}$, and the eigenvalues of $\bar T_{Z,G_D,L_y,d}$ that
lack a factor of $w$ as $\bar \lambda_{Z,G_D,L_y,d,j}$.

Furthermore, the common eigenvalues $\lambda_{Z,G_D,L_y,d,j}$ at level $d$ also
appear as $\lambda_{Z,G_D,L_y,d+1,j}$ at level $d+1$ with $0 \le d \le
L_y$. This again is due to the fact that $V_{Z,G_D,L_y,d}$ does not include the
factor $(vI+D_{L_y,d,L_y+1})$, so that it does not matter if the $(L_y+1)$-th
vertex is assigned a color (connected to a black circle) or not. Denote the
number of these common eigenvalues as $n_{Zh}(G_D,L_y,d)$ with $1 \le d \le
L_y+1$. They are given by
\beq
n_{Zh}(G_D,L_y,d) = n_{Zh}(L_y,d-1) + n_{Zh}(L_y,d)
\label{nzgd}
\eeq
That is, they can be classified as being in either category (a) plus category
(c) in the proof of Theorem \ref{nzlyd}, or the circle case in category (b)
plus category (d) with $d$ replaced by $d-1$ in that proof.  In Table
\ref{ntctablegd} we list the first few numbers $n_{Zh}(G_D,L_y,d)$ and their
total sums $N_{Zh,G_D,L_y}$.  In particular, the numbers $n_{Zh}(G_D,L_y,1)$ is
\beq
n_{Zh}(G_D,L_y,1) = \sum_{k=1}^{L_y+1} {L_y \choose k-1} C_k
\label{nzlygdd1}
\eeq
\beq
n_{Zh}(G_D,L_y,L_y)=3L_y \ , \quad n_{Zh}(G_D,L_y,L_y+1)=1
\label{nzgdvalues}
\eeq
If we denote the generating function of $n_{Zh}(G_D,L_y,1)$ as 
\beq
A_1(x) = \sum_{L_y=0}^\infty n_{Zh}(G_D,L_y,1) x^{L_y+1} = 
x+3x^2+10x^3+36x^4+...
\eeq
then the generating function of $n_{Zh}(G_D,L_y,2)$, denoted as $A_2(x)$, is 
given by the convolution \cite{bs}
\beq
A_2(x) = \sum_{L_y=1}^\infty n_{Zh}(G_D,L_y,2) x^{L_y+1} = [A_1(x)]^2 = 
x^2+6x^3+29x^4+132x^5+...
\eeq
In general, the generating function of $n_{Zh}(G_D,L_y,d)$ is given as
\beq
A_d(x) = \sum_{L_y=d-1}^\infty n_{Zh}(G_D,L_y,d) x^{L_y+1} = [A_1(x)]^d 
\eeq
From Eq. (\ref{nzgd}), it is clear that 
\beq
N_{Zh,G_D,L_y} = 2N_{Zh,L_y} - n_{Zh}(L_y,0)
\eeq
It follows that  
\beq
N_{Zh,G_D,L_y+1} = 5N_{Zh,G_D,L_y} - n_{Zh}(G_D,L_y,1) 
\label{nttotgdrecursion}
\eeq
similar to Eq. (\ref{nttotrecursion}), and $N_{Zh,G_D,L_y}$ can be expressed 
as 
\beq
N_{Zh,G_D,L_y} = \sum_{j=0}^{L_y} {L_y \choose j} {2j+1 \choose j+1} 
\eeq

\begin{table}
\caption{\footnotesize{Table of numbers $n_{Zh}(L_y,G_D,d)$ and their sums, 
$N_{Zh,G_D,L_y}$. Blank entries are zero. }}
\begin{center}
\begin{tabular}{|c|c|c|c|c|c|c|c|c|c|c|c|}
\hline\hline 
$L_y  \ \backslash \ d$ 
   & 1     & 2     & 3    & 4     & 5    & 6   & 7   & 8  & 9 &10 & 
$N_{Zh,G_D,L_y}$ \\ \hline\hline
1  & 3     & 1     &      &       &      &     &     &    &   &   & 4       \\ \hline
2  & 10    & 6     & 1    &       &      &     &     &    &   &   & 17      \\ \hline
3  & 36    & 29    & 9    & 1     &      &     &     &    &   &   & 75      \\ \hline
4  & 137   & 132   & 57   & 12    & 1    &     &     &    &   &   & 339     \\ \hline
5  & 543   & 590   & 315  & 94    & 15   & 1   &     &    &   &   & 1558    \\ \hline
6  & 2219  & 2628  & 1629 & 612   & 140  & 18  & 1   &    &   &   & 7247    \\ \hline
7  & 9285  & 11732 & 8127 & 3605  & 1050 & 195 & 21  & 1  &   &   & 34016   \\ \hline
8  & 39587 & 52608 & 39718& 19992 & 6950 & 1656& 259 & 24 & 1 &   & 160795  \\ \hline
9  & 171369& 237129&191754& 106644& 42498&12177& 2457& 332& 27& 1 & 764388  \\ \hline\hline  
\end{tabular}
\end{center}
\label{ntctablegd}
\end{table}

The multiplicity for the common eigenvalues with $1 \le d$ is defined as
\beq
\tilde \kappa^{(d)}(q) = \kappa^{(d)}(q-1) = 
\tilde c^{(d)} + \tilde c^{(d-1)} =
\sum_{j=0}^{d-1} (-1)^j { 2d-1-j \choose j} (q-1)^{d-j}
\label{barkappa}
\eeq
where $\kappa^{(d)}(q)$ is given in Eq. (\ref{kappad}). The first few of these
coefficients are
\beqs
& & \tilde \kappa^{(1)} = q-1 \ , \quad \tilde \kappa^{(2)} = (q-1)(q-3) \ , 
\cr\cr
& & \tilde \kappa^{(3)} = (q-1)(q-2)(q-4) \ , \quad \tilde \kappa^{(4)} = 
(q-1)(q-3)(q^2-6q+7) 
\label{kappabarn}
\eeqs
Collecting the common eigenvalues, we find that the partition function in
Eq. (\ref{zsdgtran}) can be rewritten as
\beqs 
Z(G_D, L_y \times L_x,q,v,w) & = & \sum_{d=1}^{L_y+1} \tilde \kappa^{(d)}
\sum_{j=1}^{n_{Zh}(L_y,G_D,d)} (\lambda_{Z,G_D,L_y,d,j})^m \cr\cr & & + w
\sum_{d=0}^{L_y} \tilde c^{(d)} \sum_{j=1}^{n_{Zh}(L_y,d)} (\bar
\lambda_{Z,G_D,L_y,d,j})^m
\label{zsdgtran2}
\eeqs
where $n_{Zh}(L_y,d)$ is given in Theorem \ref{nzlyd}.

\subsection{Determinants}

We find
\beqs 
det(T_{Z,G_D,L_y,d}) & = & (v^{L_y})^{n_{Zh}(L_y+1,d)} \biggl [ w^{L_y+1}
\Bigl ( 1+\frac{q}{v} \Bigr )^{L_y} (1+v)^{L_y} \biggr ]^{n_{Zh}(L_y,d)} \cr\cr
det(\bar T_{Z,G_D,L_y,d}) & = & (v^{L_y})^{n_{Zh}(L_y+1,d)} \biggl [ w \Bigl (
1+\frac{q}{v} \Bigr ) (1+v) \biggr ]^{L_y n_{Zh}(L_y,d)}
\label{detTZgdld}
\eeqs
for the full range $0 \le d \le L_y+1$.  Comparing $V_{Z,sq,L_y+1,d}$ in
Eq. (\ref{HVmatrix}) and $V_{Z,G_D,L_y,d}$ in Eq. (\ref{HVmatrixGD}), one sees
that $V_{Z,G_D,L_y,d}$ does not include the factor $(vI+D_{L_y,d,L_y+1})$ for
the self-dual square lattice, so that the power of $v$ and $(1+q/v)$ has the
factor $L_y$ rather than $L_y+1$. Comparing $H_{Z,G_D,L_y,d}$ and $\bar
H_{Z,G_D,L_y,d}$ in Eq. (\ref{HVmatrixGD}), one sees that the power of
$w$ for $\bar H_{Z,G_D,L_y,d}$ has the factor $L_y$ rather than $L_y+1$ in
$H_{Z,G_D,L_y,d}$.

Taking into account that the multiplicity of each $\lambda_{Z,G_D,L_y,d,j}$ is
$\tilde c^{(d)}$, it follows that the total determinant is
\beq
det(T_{Z,G_D,L_y}) \equiv \prod_{d=0}^{L_y+1} [det(T_{Z,G_D,L_y,d})]^{\tilde
c^{(d)}} = [v^{L_y}]^{q^{L_y+1}} \biggl [ w^{L_y+1} \Bigl ( 1+\frac{q}{v} \Bigr
)^{L_y} (1+v)^{L_y} \biggr ]^{q^{L_y}}
\label{detTZgd}
\eeq

\subsection{Eigenvalue for $d=L_y+1$, $\Lambda=G_D$}

For $d=L_y+1$, each factor of $V_{Z,G_D,L_y,d}$ in Eq. (\ref{HVmatrixGD})
reduces to a scalar $v$, and $H_{Z,G_D,L_y,d}$ and $\bar H_{Z,G_D,L_y,d}$
become one. Both of the transfer matrices $T_{Z,G_D,L_y,d}$ and $\bar
T_{Z,G_D,L_y,d}$ for $d=L_y+1$ reduce to a scalar, namely
$\lambda_{Z,G_D,L_y,L_y+1} = v^{L_y}$.

\subsection{Transfer Matrix for $d=L_y$, $\Lambda=G_D$}

The transfer matrix $T_{Z,G_D,L_y,L_y}$ has dimension $3L_y+2$. We obtain the
following general formulas:
\beq
(T_{Z,G_D,L_y,L_y})_{1,1} = v^{L_y} \ , \ (T_{Z,G_D,L_y,L_y})_{2,2} = v^{L_y}w
\ , \ (T_{Z,G_D,L_y,L_y})_{2L_y+1,2L_y+1} = v^{L_y-1}(2v+q-1)
\label{TTsdgx1}
\eeq
\beq
(T_{Z,G_D,L_y,L_y})_{2j-1,2j-1} = v^{L_y-1}(3v+q-1) \quad {\rm for} \quad
L_y \ge 2 \ \ {\rm and} \ \ 2 \le j \le L_y
\label{TTsdgx2}
\eeq
\beq
(T_{Z,G_D,L_y,L_y})_{2j,2j} = v^{L_y-1}w(1+v) \quad {\rm for} \quad
2 \le j \le L_y+1
\label{TTsdgx3}
\eeq
\beq
(T_{Z,G_D,L_y,L_y})_{2L_y+2,2L_y+1} = v^{L_y-1}(v+q-1) 
\label{TTsdgx4}
\eeq
\beq
(T_{Z,G_D,L_y,L_y})_{2j,2j-1} = v^{L_y-1}(2v+q-1) \quad {\rm for} \quad
L_y \ge 2 \ \ {\rm and} \ \ 2 \le j \le L_y
\label{TTsdgx5}
\eeq
\beq
(T_{Z,G_D,L_y,L_y})_{2j-1,2j} = v^{L_y-1}w \quad {\rm for} \quad
2 \le j \le L_y+1
\label{TTsdgx6}
\eeq
\beq
(T_{Z,G_D,L_y,L_y})_{2j-1,2j+1} = (T_{Z,G_D,L_y,L_y-1})_{2j,2j+1} = v^{L_y} \quad {\rm for} \quad 2 \le j \le L_y
\eeq
\beq
(T_{Z,G_D,L_y,L_y})_{2j+1,2j-1} = (T_{Z,G_D,L_y,L_y-1})_{2j+2,2j-1} = v^{L_y} \quad {\rm for} \quad 1 \le j \le L_y
\label{TTsdg1a}
\eeq
\beq
(T_{Z,G_D,L_y,L_y})_{j,j} = v^{L_y}(1+v) \quad {\rm for} \quad 2L_y+3 \le j \le 3L_y+2
\label{TTsdgy}
\eeq
\beqs
& & (T_{Z,G_D,L_y,L_y})_{3,2L_y+3} = (T_{Z,G_D,L_y,L_y-1})_{4,2L_y+3} \cr\cr
& = & (T_{Z,G_D,L_y,L_y})_{2j-1,2L_y+2+j} = (T_{Z,G_D,L_y,L_y-1})_{2j,2L_y+2+j} \cr\cr
& = & (T_{Z,G_D,L_y,L_y})_{2j+1,2L_y+2+j} = (T_{Z,G_D,L_y,L_y-1})_{2j+2,2L_y+2+j} = v^{L_y-1}(1+v) \quad {\rm for} \quad 2 \le j \le L_y \cr & &
\label{TTsdgwa}
\eeqs
\beq
(T_{Z,G_D,L_y,L_y})_{2L_y+2+j,2j-1} = (T_{Z,G_D,L_y,L_y-1})_{2L_y+2+j,2j+1}= v^{L_y+1} \quad {\rm for} \quad 1 \le j \le L_y
\label{TTsdgva}
\eeq
with all other elements equal to zero. $\bar T_{Z,G_D,L_y,L_y}$ is the same as $T_{Z,G_D,L_y,L_y}$ except for one element $(\bar T_{Z,G_D,L_y,L_y})_{2,2} = v^{L_y}$.

We illustrate these general formulas for the cases $L_y=1$ and $L_y=2$:
\beq
T_{Z,G_D,1,1} = \left( \begin{array}{ccccc}
v   & 0    & 0   & 0    & 0   \\
0   & wv   & 0   & 0    & 0   \\    
v   & 0    & x_2 & w    & v_1 \\
v   & 0    & x_1 & wv_1 & v_1 \\       
v^2 & 0    & v^2 & 0    & vv_1 
\end{array} \right )
\label{TTsdg11}
\eeq
\beq
T_{Z,G_D,2,2} = v \left( \begin{array}{cccccccc}
v   & 0    &  0  & 0    & 0   & 0    & 0    & 0   \\
0   & wv   &  0  & 0    & 0   & 0    & 0    & 0   \\ 
v   & 0    & x_3 & w    & v   & 0    & v_1  & v_1 \\
v   & 0    & x_2 & wv_1 & v   & 0    & v_1  & v_1 \\ 
0   & 0    & v   & 0    & x_2 & w    & 0    & v_1 \\
0   & 0    & v   & 0    & x_1 & wv_1 & 0    & v_1 \\ 
v^2 & 0    & v^2 & 0    & 0   & 0    & vv_1 & 0   \\
0   & 0    & v^2 & 0    & v^2 & 0    & 0    & vv_1     \end{array} \right ) 
\label{TTsdg22}
\eeq
Thus, in general, the matrix $T_{Z,G_D,L_y,L_y}$ is almost the same as
$T_{Z,sq,L_y+1,L_y}$ except for the first two rows. It is obvious that
$T_{Z,G_D,L_y,L_y}$ always has one eigenvalue equal to $v^{L_y}$ and one equal
to $v^{L_y}w$, and $\bar T_{Z,G_D,L_y,L_y}$ always has one eigenvalue equal to
$v^{L_y}$ with multiplicity two.

As corollaries of our general result for $T_{Z,G_D,L_y,L_y}$ we calculate the
determinant and trace:
\beqs
det(T_{Z,G_D,L_y,L_y}) & = & (v^{L_y})^{3L_y+2} \biggl [ w^{L_y+1} \Bigl (
1+\frac{q}{v} \Bigr )^{L_y} (1+v)^{L_y} \biggr ] \cr\cr det(\bar
T_{Z,G_D,L_y,L_y}) & = & (v^{L_y})^{3L_y+2} \biggl [ w \Bigl ( 1+\frac{q}{v}
\Bigr ) (1+v) \biggr ]^{L_y}
\label{detTsdglyly}
\eeqs
which is the $d=L_y$ special case of (\ref{detTZgdld}), and 
\beqs
Tr(T_{Z,G_D,L_y,L_y}) & = & 
v^{L_y-1} \Bigl [ L_y(q-1+v^2+4v) + w(v+L_y+vL_y) \Bigl ] \cr\cr 
Tr(\bar T_{Z,G_D,L_y,L_y}) & = & 
v^{L_y-1} \Bigl [ L_y(q-1+v^2+4v) + v + wL_y(1+v) \Bigl ]
\label{traceTsdglyly}
\eeqs

\section{Some Illustrative Calculations}

\subsection{ $L_{\lowercase{y}}=1$}

We give some explicit results of the transfer matrices for cyclic strips in
this section, beginning with the case $L_y=1$. We have exhibited the transfer
matrix $T_{Z,sq,1,0}$ in Eq. (\ref{TTsq10}), and the quantity
$T_{Z,sq,1,1}=v$ has been given above.  The Potts model partition function
$Z(sq,L_y \times m,cyc.,q,v,w)$ was calculated for the the circuit graph 
in \cite{GU}.  The results for eigenvalues agree.  (The actual transfer
matrices themselves are basis-dependent, and the basis used in \cite{GU} was
different from ours, so the matrices are different, but the only part of the
transfer matrices that enters into the partition function is the (powers of
the) eigenvalues.) 

\subsection{Square-Lattice Strip, $L_{\lowercase{y}}=2$}

The partition function for this case is given by the $L_y=2$ special case of 
Eq. (\ref{zgsum_transfer}).  The transfer matrix for $d=0$ is 
\beq
T_{Z,sq,2,0} = \left( \begin{array}{ccccc}
(q-1)^2+3(q-1)v+3v^2 & x_1w      & x_1w      & w^2v_1   & x_2v_1   \\
(q-1)^2+2(q-1)v+v^2  & x_1wv_1   & (q-1)w    & w^2v_1^2 & x_1v_1   \\
(q-1)^2+2(q-1)v+v^2  & (q-1)w    & x_1v_1w   & w^2v_1^2 & x_1v_1   \\
(q-1)^2+(q-1)v       & (q-1)wv_1 & (q-1)wv_1 & w^2v_1^3 & (q-1)v_1 \\
v^3                  & 0         & 0         & 0        & v^2v_1 
\end{array} \right )
\label{TZsq20}
\eeq
and the matrices $T_{Z,sq,2,1}$ and $T_{Z,sq,2,2}$ have been given above.  The
result for the cyclic case $Z(sq,2 \times m,cyc.,q,v,w)$ agrees with
Ref. \cite{mirza}.  The partition function for the M\"obius case follows from
our general formulas also given above.

\subsection{Triangular-Lattice Strip, $L_{\lowercase{y}}=2$}

We illustrate our results for the $L_y=2$ cyclic strip of the triangular
lattice. We obtain 
\beq
T_{Z,tri,2,0} = \left( \begin{array}{ccccc}
(q-1)^2+4(q-1)v+5v^2+v^3 & x_1w        & x_2w      & w^2v_1   & y_3v_1   \\
(q-1)^2+3(q-1)v+3v^2+v^3 & x_1wv_1     & x_1w      & w^2v_1^2 & y_2v_1   \\
(q-1)^2+2(q-1)v+v^2      & (q-1)wv_1   & x_1v_1w   & w^2v_1^3 & x_1v_1   \\
(q-1)^2+(q-1)v           & (q-1)wv_1^2 & (q-1)wv_1 & w^2v_1^4 & (q-1)v_1 \\
v^2y_3                   & 0           & wv^2      & 0        & v^2v_1v_2 
\end{array} \right )
\label{TZtri20}
\eeq
The matrices $T_{Z,tri,2,1}$ and $T_{Z,tri,2,2}$ were given above.

\subsection{Honeycomb-lattice Strip, $L_{\lowercase{y}}=2$}

For the $L_y=2$ cyclic strip of the honeycomb lattice we calculate
\beq
T_{Z,hc,2,0}= \left( \begin{array}{ccccc}
x_1^2    & x_1w      & x_1w      & w^2      & x_2 \\
(q-1)x_1 & x_1wv_1   & (q-1)w    & w^2v_1   & x_1 \\
(q-1)x_1 & (q-1)w    & x_1v_1w   & w^2v_1   & x_1 \\
(q-1)^2  & (q-1)wv_1 & (q-1)wv_1 & w^2v_1^2 & q-1 \\
0        & 0         & 0         & 0        & v^2 
\end{array} \right ) T_{Z,sq,2,0}
\label{TZhc20}
\eeq
where $T_{Z,sq,2,0}$ is given in Eq. (\ref{TZsq20}). The other matrices
relevant for this strip were given above.

\bigskip

Acknowledgments: We thank F. Y. Wu for a valuable communication calling our
attention to Ref. \cite{wu78}. The research of R.S. was partially supported by
the NSF grant PHY-00-98527.  The research of S.C.C. was partially supported by
the Taiwan NSC grant NSC-97-2112-M-006-007-MY3 and NSC-98-2119-M-002-001.



\vfill
\eject
\end{document}